\documentclass[prb,aps,floats,twocolumn]{revtex4}

\usepackage{amsmath}
\usepackage{mathtools}
\usepackage{amsfonts}
\usepackage{amssymb}
\usepackage{graphicx}
\usepackage{multirow}

\usepackage{color}

\begin{document}



\title{Surface states in topological semimetal slab geometries}

\author{Enrique Benito-Mat\'{\i}as}
\affiliation{Instituto de Estructura de la Materia, IEM-CSIC, Serrano 123, Madrid 28006,
Spain}
\affiliation{Departamento de Econom\'{\i}a Financiera y Contabilidad e Idioma Moderno, Universidad Rey Juan Carlos, 
28933 M\'ostoles, Spain}
\author{Rafael A. Molina}
\affiliation{Instituto de Estructura de la Materia, IEM-CSIC, Serrano 123, Madrid 28006, Spain}



\begin{abstract}
	Weyl semimetals are topological materials with protected Weyl nodes in the band structure. In these materials the surface states form open curves at the Fermi surface, Fermi arcs in Weyl semimetals and drumhead states of nodal-line semimetals.
	In this work we solve analitically the wave function of the surface states in a generic continuous model describing
	Weyl and nodal-line type I-II semimetals within a slab geometry. 
	Depending on the values of the parameters, different types of Fermi arcs and drumhead states appear. When the mass terms are dominant with respect to the Fermi velocity in the Hamiltonian the decay of the surface states become oscillatory. This property has important consequences in the stability of surface states in a slab geometry.
	This exact solution can be used for a better understanding of the behaviour of Fermi Arcs in real materials and their influence in transport and optical properties. We use these solutions to study the Joint Density of States at the surface which can be used to interpret quasi-particle interference data in scanning tunnneling microscope experiments. We show that oscillatory decay can be distinguish from simple exponential decay of the surface states in these experiments.
\end{abstract}

\maketitle

\section{Introduction}
\label{sec:intro}

Topological semimetals have attracted a lot of attention recently as they show new macroscopic quantum phenomena that, besides being of great fundamental interest, hold a lot of potential for technological applications\cite{Burkov2016}. The most studied in this family of materials are probably the Weyl semimetals which possess isolated Weyl nodes in the band structure \cite{Xu2015a}. Weyl fermions were originally considered in massless quantum electrodynamics but has not been observed as a fundamental particle. However, they can be observed as quasi-particles in such condensed-matter realizations.
The surface states corresponding to these topological materials lie on contours which do not form closed curves. In the case of Weyl semimetals, the surface states form the so-called Fermi arcs, that join on the projection of the nodes onto the given surface. Such states are topologically protected and Chern numbers can be defined in the planes lying between
Weyl nodes as they can be considered as monopoles of the Berry curvature\cite{Wan2011,Yang2011}. The physical quantity measuring this protection and the robustness of the Weyl node structure to perturbations is the separation of the nodes in momentum space. Dirac semimetals like Na$_3$Bi and Cd$_3$As$_2$ can be viewed as Z$_2$ Weyl semimetals where the chiral structure of the nodes is protected by particle-hole symmetry \cite{Gorbar2015b}. Other important members of this family of materials are the nodal line semimetals which instead of isolated Weyl nodes in the bulk present a continuous symmetry-protected line of nodes \cite{Burkov2016}. The surface states in this case form 2D manifolds and are called drumhead states due to their shape in the Brillouin zone \cite{Fang2016}.


The presence of quantum anomalies is one of the most important properties of topological semimetals. An anomaly in quantum field theory is the breaking of a classically allowed symmetry by quantum effects. The chiral (or Adler-Bell-Jackiw) anomaly first appeared
in pion decay as the coupling to the electromagnetic field breaks the chiral symmetry of pions \cite{Jackiw2008}. In Weyl semimetals the chiral anomaly manifests as a large negative longitudinal magnetoresistance due to charge being pumped between Weyl nodes at a rate proportional to the scalar product between external electric and magnetic fields \cite{Nielsen1983,Wan2011,Burkov2011}. Large negative magnetorresitance have been measured in Dirac semimetals \cite{Liang2015} and in Weyl semimetals \cite{Shekhar2015}. 
The lesser symmetry in condensed-matter as opposed to fundamental particles allow for type II Weyl semimetals where there are hole pockets at the same energy as the Weyl nodes. As the density of bulk states at the Fermi energy is larger in type II Weyl semimetals, they present different properties from the standard type I Weyl semimetals\cite{Soluyanov2015}. This classification can be extended to nodal-line semimetals \cite{Zhang2017}.

Surface states of topological semimetals have been addressed in Angle-Resolved-Photoemission Spectroscopy (ARPES) experiments which have provided the main evidence for the existence of Fermi arcs \cite{Xu2015a,Lv2015a,Lv2015b,Yang2015,Xu2015b}. However, ARPES has resolution limitations, only probes occupied states and cannot be used in the presence of a magnetic field. Surface states have also been studied through real space imaging techniques from Scanning Tunneling Microscopy (STM) \cite{Zheng2018}. Impurities scatter the surface electrons and produce a standing wave pattern on the surface which depends on the surface momenta of the electrons at the tunneling energy probed. The resulting signal is called Quasi-Particle Interference (QPI), has a high energy resolution, can be used in the presence of a magnetic field and is not limited to occupied states \cite{Hoffman2002}. Fourier transforming the QPI pattern maps the momentum transfer in the surface state. Although, the properties of the particular impurity will greatly influence the result, a simple computation of the Joint density of states (JDOS) for the surface states can help to interpret the experimental QPI patterns \cite{Kourtis2016}. With the analytic formulas for the dispersion relation and existence domains in momentum space, this can be done in a very simple way. However, quantum interference may induce a strong supression of intra-arc scattering which is an effect not captured in the JDOS autocorrelation \cite{Mitchell2016}. QPI measurement have already been used to investigate surface states of topological materials. In topological insulators, they beautifully show the absence of backscattering by normal impurities but not by magnetic ones \cite{Roushan2009,Sessi2014}. Surface states of type I Weyl semimetals from the family of TaAs have also been investigated with QPI \cite{Inoue2016,Batabyale2016} as well as type II Weyl semimetals from the WTe$_2$ family \cite{Zheng2016}.



In recent experiments thin films of the three-dimensional Dirac semimetal $Cd_3As_2$ have been grown by Molecular Beam Epitaxy\cite{Schumann2018}. The observation of the quantum Hall effect in these confined structures show that, in sufficiently thin films and at low temperatures, surface states dominates electric transport. Thus, it would be desirable to have analytic solutions in this type of geometry with easily interpretable physical properties.
Although, solutions for the semi-infinite system of Weyl semimetals have already been obtained \cite{Gorbar2015a,Gonzalez2017}, only a partial picture of the Hilbert space for these surface states has been achieved. In this work we derive a complete map for the surface states in slab geometries, providing analytical and explicit formulas for low energy continuous models describing Weyl, Dirac and Nodal line semimetals.

The work is structured as follows: In section \ref{sec: Model_Hamiltonian} we introduce the models. Section \ref{sec: General_solution} provides the notation and the general solutions without boundary conditions. In Section \ref{sec:Fermi_arcs_in_a_slab} we solve the problem within the slab and compute the analytical formulas.
Then we present a study of the limiting behavior for a thick slab that provides a reference frame to compare with the slab solution where we will see that some of the states for the thick slab survive in quantized domains. JDOS diagrams are computed for several cases of interest. Finally, in section \ref{sec:conclusions} we present the conclusions. The appendix contains most of the algebraic manipulations.




\section{Model Hamiltonian}
\label{sec: Model_Hamiltonian}

We study a simple model for a Weyl semimetal with two Weyl nodes considering terms up to quadratic order in the quasimomentum.
\begin{eqnarray}
H=\varepsilon_0(\textbf{k})I+M(\textbf{k})\sigma_z+\upsilon(\zeta k_x \sigma_x-k_y \sigma_y).
\label{eqn:weylhamiltonian}
\end{eqnarray}
\begin{eqnarray*}
\varepsilon_0(\textbf{k})=c_0 + c_1k_z^2+c_2(k_x^2+k_y^2),
\end{eqnarray*}
\begin{eqnarray*}
M(\textbf{k})=m_0-m_1k_z^2-m_2(k_x^2+k_y^2),
\end{eqnarray*}
where $I$ stands for the 2$\times$2 identity matrix, $\sigma_i, i=x, y, z$ are the Pauli matrices and $\zeta=\pm 1$ sets the chirality in the Dirac cones.
The same Hamiltonian with $\zeta=0$ can describe a nodal-line semimetal \cite{Burkov2016}.

\begin{figure}
	\includegraphics[width=0.5\textwidth]{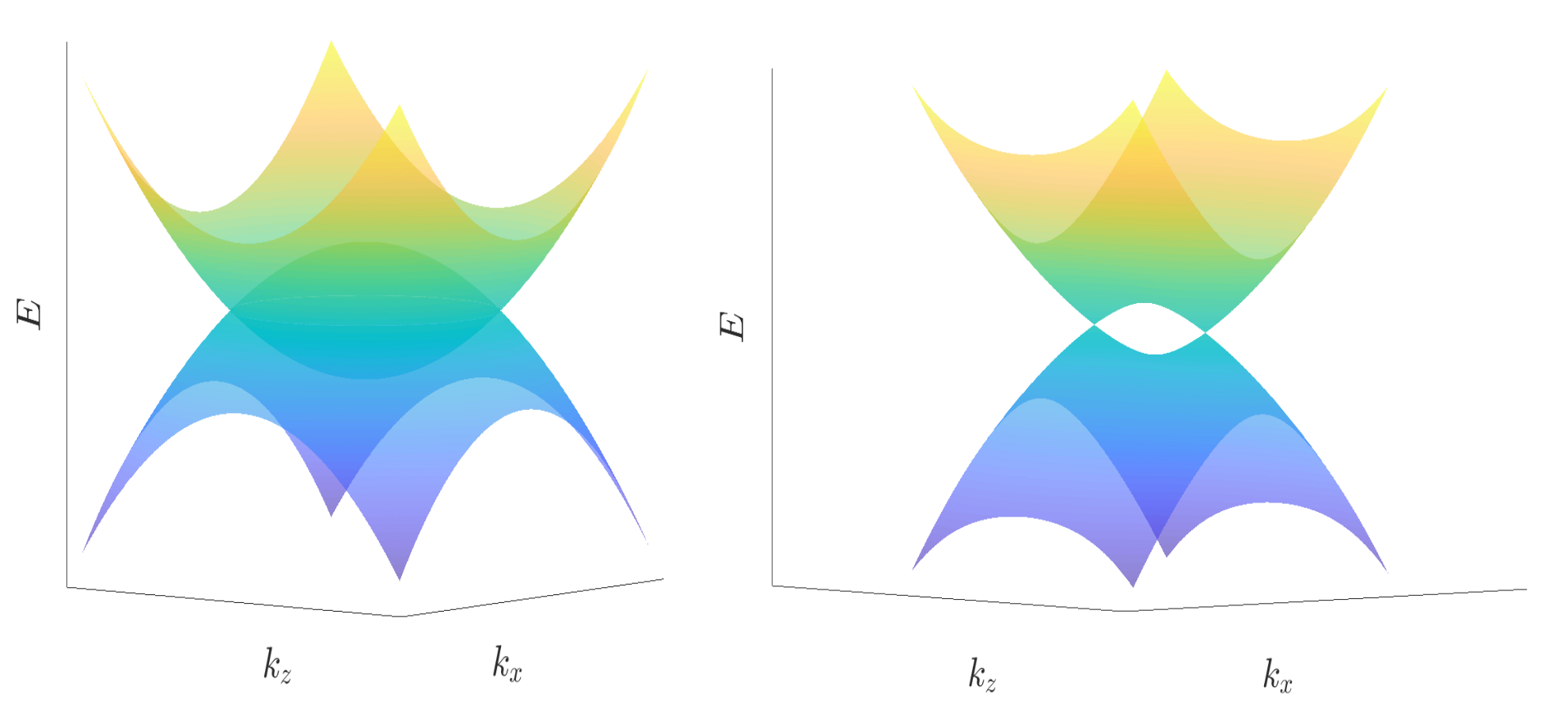}
	\caption{\label{fig:weylnodes}(Right) Band structure around the Weyl nodes for the Hamiltonian (\ref{eqn:weylhamiltonian}) with model parameters $c_0=c_1=c_2=0$, $m_0=-0.1, m_1=m_2=-1, \upsilon=1$ and $k_y=0$. (Left) Nodal-line semimetal ($\zeta=0$) for the same paremeters}
\end{figure}

This Hamiltonian has been proposed, for example, as a low-energy description of the {\em ab initio} DFT results for the family of compounds A$_3$ Bi (A=Na, K, Rb) \cite{Wang2012,Wang2013} which are actually $\mathbb{Z}_2$ Weyl semimetals \cite{Gorbar2015a}. The eigenvalues of the Hamiltonian are
\begin{equation}
E(\textbf{k})=\epsilon_0(\textbf{k})\pm \sqrt{M^2(\textbf{k})+\upsilon^2(\zeta^2k_x^2+k_y^2)}.
\label{eqn:dispersion}
\end{equation}
For $\zeta=\pm1$, there are two Weyl points (Fig.\ref{fig:weylnodes}: Right) at momentum positions, $\textbf{k}_0^{\pm}=(0,0,\pm\sqrt{m_0/m_1})$. If we consider the $4 \times 4$ matrix with both values of $\zeta$, the Weyl points transform into Dirac points with both degenerate chiralities in the same node. However, the topological properties and Fermi arcs remain the same as for the $2 \times 2$ model with non-degenerate Weyl nodes as they are protected by up-down parity symmetry\cite{Gorbar2015a}. For $\zeta=0$ there is a continous line of nodes in the plane $k_y=0$ given by the elliptical set 
$m_2k_x^2+m_1k_z^2=m_0$ (Fig. \ref{fig:weylnodes}: Left). Depending on the values of $\epsilon_0(\textbf{k})$ the nodes may be more or less tilted and be a type I or II Weyl semimetal (or nodal-line semimetal). Specifically, the transition to a type II semimetal occurs in this model for $c_1^2>m_1^2$

\section{General solution}
\label{sec: General_solution}








In order to solve the eigenvalue problem with the appropiate boundary conditions, we use an \textit{ansatz} wave function with the following structure:

\begin{align}
\psi(\textbf{r})_\zeta=f(x,z)_b\psi_{k_x, k_z}(y)_\zeta
\end{align}

where, without loss of generality, $f(x,z)_b$ is a free wave function including all normalization constants, and

\begin{align}
\psi_{k_x, k_z}(y)_\zeta=\sum_i^N A_{i}e^{-\lambda_iy}(\Phi_{\lambda_i, \zeta}),
\label{eqn:ansatz1}
\end{align}

with $\Phi_{\lambda_i, \zeta}$ being position independent (y-independent) spinors.
Since the system of Differential equations is linear, all the $A_{\lambda_i}e^{\lambda_iy}(\Phi_{\lambda_i, \zeta})$ must be a solution to the Schr\"odinger equation corresponding to the Hamiltonian (\ref{eqn:weylhamiltonian}) but, only the correct linear combination will fulfill the appropiate boundary conditions.

Then, trying the \textit{anstaz} function $e^{\lambda y}(\Phi_{\lambda_i, \zeta})$
 in the Schr\"odinger equation we obtain the following eigenvalue problem:
\begin{widetext}
\begin{align}
\begin{pmatrix}
c_{2-}(-\lambda^2+k_x^2)+\theta_- -E &  \upsilon(\zeta k_x-\lambda)\\ \upsilon(\zeta k_x+\lambda) &  c_{2+}(-\lambda^2+k_x^2)+\theta_+ -E
\end{pmatrix} \begin{pmatrix} \Phi_1\\\Phi_2
\end{pmatrix}_{\lambda, \zeta}=\begin{pmatrix} 0\\0\end{pmatrix}
\label{eqn:weylhamiltonian5}
\end{align}
\end{widetext}

Where $\Phi_{\lambda, \zeta}=\begin{pmatrix} \Phi_1\\\Phi_2
\end{pmatrix}_{\lambda, \zeta}$ is the spinor, $c_{2\pm}=c_2\pm m_2$ and $\theta_{\pm}=c_0\mp m_0+(c_1 \pm m_1)k_z^2$.


This determines a biquadratic equation in $\lambda$ with four roots for a given energy: 

\begin{equation}
\label{eqn:roots}
\begin{aligned}
\lambda_{3,4}=-\sqrt{k_x^2-\frac{1}{2c_{2+}c_{2-}}(-b\pm\sqrt{b^2-Q_{W/N}^2})}\\
\lambda_{1,2}=-\lambda_{3,4}
\end{aligned}
\end{equation}

With:

\begin{equation}
\begin{cases}b=c_{2+}(\theta_- -E)+c_{2-}(\theta_+ -E)-\upsilon^2\\
Q_W^2=4c_{2+}c_{2-}(\theta_- -E)(\theta_+ -E) & Weyl\\
Q_N^2=4c_{2+}c_{2-}[(\theta_- -E)(\theta_+ -E)+k_x^2\upsilon^2] & Nodal\end{cases}
\end{equation}

Depending on the model considered (Weyl or nodal line semimetal). Since $E$ must be real there are some restrictions over the possible values for $\lambda_{1,2}$ (see \ref{sec:appendix}). The values of the different $\lambda$'s can be purely imaginary but these are bulk solutions that do not concern us here. The surface states can have purely real values of $\lambda$ or they can have both real and imaginary parts.
In a much simpler model for Weyl and nodal line semimetals with a reduced number of parameters the real case has been named type B surface states and they decay from the surface with a purely exponential decay. The imaginary case presents oscillations on top of the exponential decay and has been named type A surface states \cite{Gonzalez2017}. We will follow this notation here. It is interesting to note that these type of states have been shown to be exceptional points upon complexification of the momentum in the Hamiltonian, turning it into a non-Hermitian Hamiltonian. This procedure has been shown to be fruitful for topological semimetals in different situations \cite{Gonzalez2016,Gonzalez2017,Molina2017}. It can be proven that this is also the case in the more complex Hamiltonian we are analyzing here (see the analysis in the next paragraph) including type I and II Weyl semimetals. However, in this work, we will not pursue those ideas any further. 
 
The two component spinor is (apart from normalization constant):

\begin{align}
\Phi_{\lambda, \zeta}=\begin{pmatrix} \upsilon(\lambda-\zeta k_x)\\c_{2-}(k_x^2-\lambda^2)+(\theta_- -E)\end{pmatrix}
\label{eqn:spinor1}
\end{align}




so, the general solution for the wave function (\ref{eqn:ansatz1}) reads:
\begin{eqnarray}
\psi_{k_y, k_z}(y)_\zeta=A_{1}\Phi_{\lambda_1, \zeta}e^{-\lambda_1y}+A_{2}\Phi_{\lambda_2, \zeta}e^{-\lambda_2y}+ \nonumber \\
A_{-1}\Phi_{-\lambda_1, \zeta}e^{\lambda_1y}
+A_{-2}\Phi_{-\lambda_2, \zeta}e^{\lambda_2y},
\label{eqn:wavef1}
\end{eqnarray}
with the different $A_{i},\, i=1,2,-1,-2$ being the different amplitudes of the linear combination that must be normalized.


Before we proceed into the slab solution, there are some subtle questions about the possible roots for these surface states that we would like to clarify here. In order to understand this properly, we have to compute the free dispersion relation (\ref{eqn:dispersion}) for the Fermi Arcs: $k_y \to i\lambda$

\begin{equation}
E=\epsilon_0\pm \sqrt{M^2+\upsilon^2(\zeta^2k_x^2-\lambda^2)}.
\label{eqn:dispersionss}
\end{equation}

\begin{figure}[htb!]
	\includegraphics[width=0.5\textwidth]{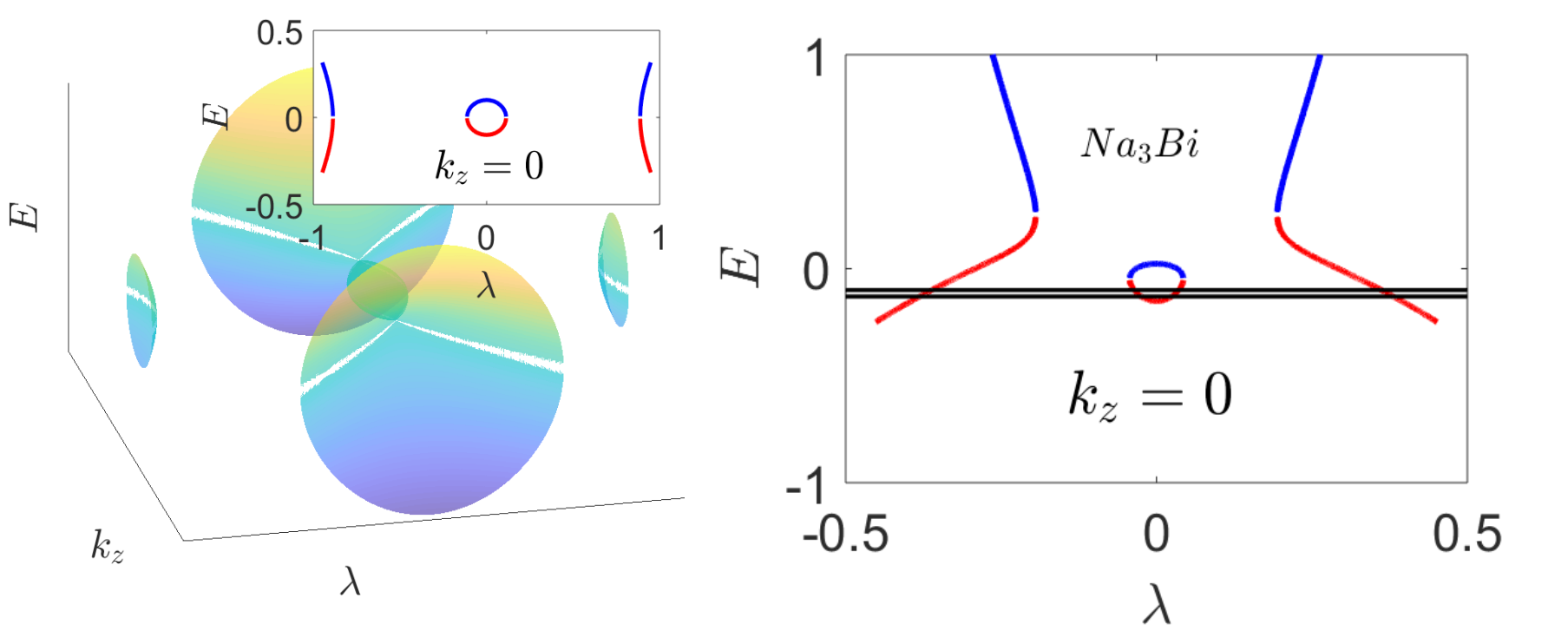}
	\caption{\label{fig:dispersionfa}(Left panel) Dispersion relation (\ref{eqn:dispersionss}) with model parameters $c_0=c_1=c_2=0$, $m_0=-0.1, m_1=m_2=-1, \upsilon=1$ and $k_x=0$ and its projection over the plane $k_z=0$ (Right panel) The same projection in the dispersion relation for the $Na_3Bi$ compound. The horizontal lines would be the energies for the two Fermi Arcs so, the intersections with the red branch gives a group of four roots for each state}
\end{figure}

The energies consisting on disjoint domains within each energy branch (see FIG. \ref{fig:dispersionfa}). Using dispersion relation (\ref{eqn:dispersionss}) we have a first estimation for the roots or the penetration depths here defined as $l=1/Re(\lambda)$:

Weyl semimetal:
\begin{widetext}
\begin{equation}
\begin{aligned}
\label{eqn:weylfirstpendepth}
\lambda_1\geqslant\sqrt{k_x^2+\frac{\upsilon^2}{2m_2^2}-\frac{(m_0-m_1k_z^2)}{m_2}+\sqrt{\left(\frac{\upsilon^2}{2m_2^2}\right)^2-\left(\frac{m_0-m_1k_z^2}{m_2^3}\right)\upsilon^2}}\Rightarrow l_1\leqslant 1/\lambda_1\\
\lambda_2\leqslant\sqrt{k_x^2+\frac{\upsilon^2}{2m_2^2}-\frac{(m_0-m_1k_z^2)}{m_2}-\sqrt{\left(\frac{\upsilon^2}{2m_2^2}\right)^2-\left(\frac{m_0-m_1k_z^2}{m_2^3}\right)\upsilon^2}}\Rightarrow l_2\geqslant 1/\lambda_2
\end{aligned}
\end{equation}
\end{widetext}
Nodal line semimetal:
\begin{widetext}
	\begin{equation}
	\begin{aligned}
	\label{eqn:nodalfirstpendepth}
	\lambda_1\geqslant\sqrt{k_x^2+\frac{\upsilon^2}{2m_2^2}-\frac{(m_0-m_1k_z^2)}{m_2}+\sqrt{\left(\frac{\upsilon^2}{2m_2^2}\right)^2+\left(\frac{m_2k_x^2-(m_0-m_1k_z^2)}{m_2^3}\right)\upsilon^2}}\Rightarrow l_1\leqslant 1/\lambda_1\\
	\lambda_2\leqslant\sqrt{k_x^2+\frac{\upsilon^2}{2m_2^2}-\frac{(m_0-m_1k_z^2)}{m_2}-\sqrt{\left(\frac{\upsilon^2}{2m_2^2}\right)^2+\left(\frac{m_2k_x^2-(m_0-m_1k_z^2)}{m_2^3}\right)\upsilon^2}}\Rightarrow l_2\geqslant 1/\lambda_2
	\end{aligned}
	\end{equation}
\end{widetext}

The estimates (\ref{eqn:weylfirstpendepth}) for the Na$_3$Bi compound give (see \ref{subsec:isolated_edge} for the values of the model parameters) $\lambda_1\geqslant 0.1943\AA{}^{-1}\Rightarrow l_1\lesssim 5\AA{}$, $\lambda_2\leqslant 0.0432\AA{}^{-1}\Rightarrow l_2\gtrsim 23\AA{}$ ($k_x=k_z=0$, Na$_3$Bi lattice constant $\sim 7.5\AA{}$). 

Now, in the case $c_2=0$, like in the toy model, the two branches are symmetric (FIG. \ref{fig:dispersionfa}: Left panel) with respect to the $E=0$ axis, but this is not true if $c_2 \neq 0$, which is the generic case for real materials. Since red there is a pair of Fermi Arcs and since, as we will see, one of the effects of the slab is to open a gap between them, the two surface states have different roots or penetration depths, corresponding to different intersections with the same branch of  (\ref{eqn:dispersionss}). (FIG. \ref{fig:dispersionfa}: Right panel).

\section{Fermi arcs in a slab}
\label{sec:Fermi_arcs_in_a_slab}
We will impose boundary conditions in a slab geometry of width $w$ such that $\psi_{k_x, k_z}(-w/2)_\zeta=\psi_{k_x, k_z}(w/2)_\zeta=0$. 

After some tedious but straightforward algebra we can arrive
to the necessary conditions through a $4 \times 4$ system of equations with the amplitudes $A_i$ as the unknown quantities:

\begin{widetext}
\begin{equation}
\begin{aligned}
A_{2}\Phi_{\lambda_2, \zeta}=A_{-1}\Phi_{-\lambda_1, \zeta}\dfrac{\text{sinh}(\lambda_1-\lambda_2)w/2}{\text{sinh}(\lambda_2w)}-A_{1}\Phi_{\lambda_1, \zeta}\dfrac{\text{sinh}(\lambda_1+\lambda_2)w/2}{\text{sinh}(\lambda_2w)}\\
A_{-2}\Phi_{-\lambda_2, \zeta}=A_{1}\Phi_{\lambda_1, \zeta}\dfrac{\text{sinh}(\lambda_1-\lambda_2)w/2}{\text{sinh}(\lambda_2w)}-A_{-1}\Phi_{-\lambda_1, \zeta}\dfrac{\text{sinh}(\lambda_1+\lambda_2)w/2}{\text{sinh}(\lambda_2w)}\\
\end{aligned}
\label{eqn:fundamentalsystem}
\end{equation}
\end{widetext}

For a non-trivial solution in the coefficients $A_1, A_2, A_{-1}, A_{-2}$ we need the last determinant to be zero and in doing so, we obtain an independent 
relation of (\ref{eqn:roots}) between the energy and the roots:

\begin{widetext}
\begin{equation}
(\lambda_1-\lambda_2)^2\Delta E(-\lambda_1,-\lambda_2)\Delta E(\lambda_1,\lambda_2)\text{sinh}^2((\lambda_1+\lambda_2)w/2)=(\lambda_1+\lambda_2)^2\Delta E(\lambda_1,-\lambda_2)\Delta E(-\lambda_1,\lambda_2)\text{sinh}^2((\lambda_1-\lambda_2)w/2)
\label{eqn:dispersionequation}
\end{equation}
\end{widetext}
With $\Delta E(\lambda_1,\lambda_2)=\theta_--E + c_{2-}[k_x^2+k_x\zeta(\lambda_1 + \lambda_2)+\lambda_1\lambda_2]$ the difference between the Energy for the slab and its limiting behavior when $w\to\infty$ (see \ref{subsec:isolated_edge}).
Then, solving for $E$ and defining $g^{\pm}=(\lambda_1\pm\lambda_2)^2\text{sinh}^2((\lambda_1\mp\lambda_2)w/2)$, $\varGamma=\dfrac{g^++g^-}{g^+-g^-}$ the two solutions are:
\begin{widetext}
\begin{equation}
E=\theta_-+c_{2-}\left[k_x^2-\lambda_1\lambda_2\varGamma\pm\sqrt{\zeta^2k_x^2(\lambda_1^2+\lambda_2^2-2\lambda_1\lambda_2\varGamma)+\lambda_1^2\lambda_2^2(\varGamma^2-1)}\right]
\label{eqn:disp3}
\end{equation}
\end{widetext}
With corresponding wave functions:

\small $\left( f(y)=e^{\lambda_1y}+\dfrac{\text{sinh}(\lambda_1-\lambda_2)w/2}{\text{sinh}(\lambda_2w)}e^{-\lambda_2y}-\dfrac{\text{sinh}(\lambda_1+\lambda_2)w/2}{\text{sinh}(\lambda_2w)}e^{\lambda_2y}\right) $
\normalsize
\begin{widetext}
	\begin{equation}
	\begin{aligned}
	\psi_{k_x, k_z}(y)_\zeta=A_{1}\left(\Phi_{\lambda_1, \zeta}f(-y)-\dfrac{\sqrt{g^-}\Delta E(-\lambda_1,-\lambda_2)}{\sqrt{g^+}\Delta E(\lambda_1,-\lambda_2)}\Phi_{-\lambda_1, \zeta}f(y)\right)
	\label{eqn:wavef2}=A_{1}\left(\Phi_{-\lambda_1, \zeta}f(y)-\dfrac{\sqrt{g^-}\Delta E(\lambda_1,\lambda_2)}{\sqrt{g^+}\Delta E(-\lambda_1,\lambda_2)}\Phi_{\lambda_1, \zeta}f(-y)\right)
	\end{aligned}
	\end{equation}
	\end{widetext}
Evaluating the Hamiltonian (\ref{eqn:weylhamiltonian5}) for the wave functions (\ref{eqn:wavef2}) at the edges $y=\pm w/2$ we can also find a relation between $\lambda_1$ and $\lambda_2$ in the slab:
\begin{equation}
2\lambda_1\lambda_2c_{2-}c_{2+}(\varGamma+1)=\upsilon^2+c_{2-}c_{2+}(\lambda_1+\lambda_2)^2
\label{trickequation}
\end{equation}
This will be very useful in the limiting behavior $w\to\infty$ that we will study in the following subsection and, although it can be derived in that case in a much simpler way, it serves to establish the correspondence between the expressions for a thick slab and an arbitrary width slab.

Equations (\ref{eqn:disp3},\ref{eqn:roots}) determine implicitly the Energy and the two roots $\lambda_{1,2}$ so, through relation (\ref{eqn:wavef2}) the wave function is then known, apart from normalization constant.
However, the Hilbert space possesses a rich structure most of which can be obtained explicitly. That is what we will show in next sections.

\subsection{Limiting behavior: Isolated edges}
\label{subsec:isolated_edge}
First of all, we study the limiting behaviour of one isolated edge. This highly simplify the problem so that all quantities can be obtained in explicit form. Moreover, it represents a reference frame to understand, precisely, the consequences of the interaction between the two edges. For some of the particular parameter ranges, the case of one isolated edge has already been solved \cite{Gorbar2015b,Gonzalez2017}. However, a complete map for the Hilbert space of Fermi Arcs is still absent, even in this simplified situation. In particular, Fermi Arcs fall into two different categories that behave differently: $\lambda_{1,2}\in\mathbb{R}$ (type B)or $\lambda_{1,2}\in\mathbb{C}, \notin \mathbb{R}$ (type A). The possibility of $\lambda_{1,2}\in\mathbb{C},\notin \mathbb{R}$ is usually ignored so its analysis is one the most important results in this work. To explore the semi-infinite system, we take the limit $w\rightarrow\infty$ in system (\ref{eqn:fundamentalsystem}) and results (\ref{eqn:disp3}, \ref{eqn:wavef2}, \ref{trickequation})
%
%
%
%
%
%
Then the possible solutions decouple between the two edges, $\varGamma\to-1\Rightarrow\lambda_1+\lambda_2=\dfrac{\upsilon}{\sqrt{-c_{2-}c_{2+}}}$, $E=\theta_- + c_{2-}[k_x^2\pm k_x\zeta(\lambda_1 + \lambda_2)+\lambda_1\lambda_2]$ (equivalent to $\Delta E(\lambda_1,\lambda_2)\Delta E(-\lambda_1,-\lambda_2)=0$ in (\ref{eqn:dispersionequation})) and the wave function (\ref{eqn:wavef2}) factorize in the sum of two exponentials by a spinor: 
$\psi_{k_x, k_z}^\pm(y)_\zeta=A_{\mp 1}\Phi_{\mp\lambda_1, \zeta}(e^{\lambda_1(\pm y-w/2)}-e^{\lambda_2(\pm y-w/2)})$. After some algebra, everything can be obtained explicitly as a function of model parameters.

The dispersion
\small
\begin{align}
E^\pm=\left(c_0+c_2\left(\dfrac{m_0}{m_2}\right)\right)+\left(c_1-c_2\left(\dfrac{m_1}{m_2}\right)\right)k_z^2\mp\zeta\upsilon\dfrac{\sqrt{m_2^2-c_2^2}}{m_2} k_x,
\label{dispertionIE3}
\end{align}
\normalsize
is linear in $k_x$ and parabolic in $k_z$.


The roots are:
\small
\begin{equation}
\begin{aligned}
\lambda_1^\pm=\Delta+\sqrt{F}=\Delta+\sqrt{(k_x\mp \zeta k_{x,0})^2+\dfrac{k_z^2}{\left(\dfrac{m_2}{m_1} \right)}+\Delta^2-R_\zeta^2}\\
\lambda_2^\pm=\Delta-\sqrt{F}=\Delta-\sqrt{(k_x\mp \zeta k_{x,0})^2+\dfrac{k_z^2}{\left(\dfrac{m_2}{m_1} \right)}+\Delta^2-R_\zeta^2}
\end{aligned}
\label{rootsIE}
\end{equation}
\normalsize

Where $\Delta=\dfrac{\upsilon}{2\sqrt{m_2^2-c_2^2}}$, $k_{x,0}=\left(\dfrac{c_2}{m_2}\right)\Delta$ and $R_\zeta^2=\left(\dfrac{m_0}{m_2}\right)+\zeta^2\Delta^2 \left(\dfrac{c_2}{m_2}\right)^2$

The superscripts $\pm$ in the roots are used here just to remain that, as was pointed out at the end of section \ref{sec: General_solution}, the roots are different for different Fermi arcs, something that is clear in this case looking at their explicit expressions (\ref{rootsIE}).
And also we can obtain the wave function with an explicit form for the spinors:
\begin{equation}
\begin{aligned}
\psi(\pm w/2)=0 \Rightarrow \psi_{k_x, k_z}^\pm(y)_\zeta=\\=A_{\mp 1}\begin{pmatrix}\mp 1\\\sqrt{\dfrac{m_2-c_2}{c_2+m_2}}\end{pmatrix}(e^{\lambda_1(\pm y-w/2)}-e^{\lambda_2(\pm y-w/2)})\\ 
\end{aligned}
\label{eqn:spinor_IE}
\end{equation}

The values of $F$ determine if the roots are real numbers or complex conjugate of each other as we see. Now, by definition $\lambda_{1,2}$ are such that: $Re(\lambda_{1,2})>0$, but this is so if and only if: $F<\Delta^2$. To be more precise:

\begin{flushleft}
$0<F<\Delta^2$
\end{flushleft}
\begin{equation}
\begin{aligned}
\lambda_{1,2}^\pm\in\mathbb{R}\Leftrightarrow
1-\dfrac{\Delta^2}{R_\zeta^2}<\dfrac{(k_x\mp \zeta k_{x,0})^2}{R_\zeta^2}+\dfrac{k_z^2}{\left(R_\zeta \sqrt{\dfrac{m_2}{m_1}}\right)^2}<1 
\end{aligned}
\label{trickresultconditionIE2}
\end{equation}
\begin{flushleft}
$F<0$ 
\end{flushleft}
\begin{equation}
\begin{aligned}
\lambda_{1,2}^\pm\in\mathbb{C}/  \lambda_1^\pm=(\lambda_2^\pm)^*\Leftrightarrow
\dfrac{(k_x\mp \zeta k_{x,0})^2}{R_\zeta^2-\Delta^2}+\dfrac{k_z^2}{(R_\zeta^2-\Delta^2) \dfrac{m_2}{m_1}}<1 
\end{aligned}
\label{trickresultconditionIE2c}
\end{equation}

\begin{figure}[htb!] 	
	\includegraphics[width=0.5\textwidth]{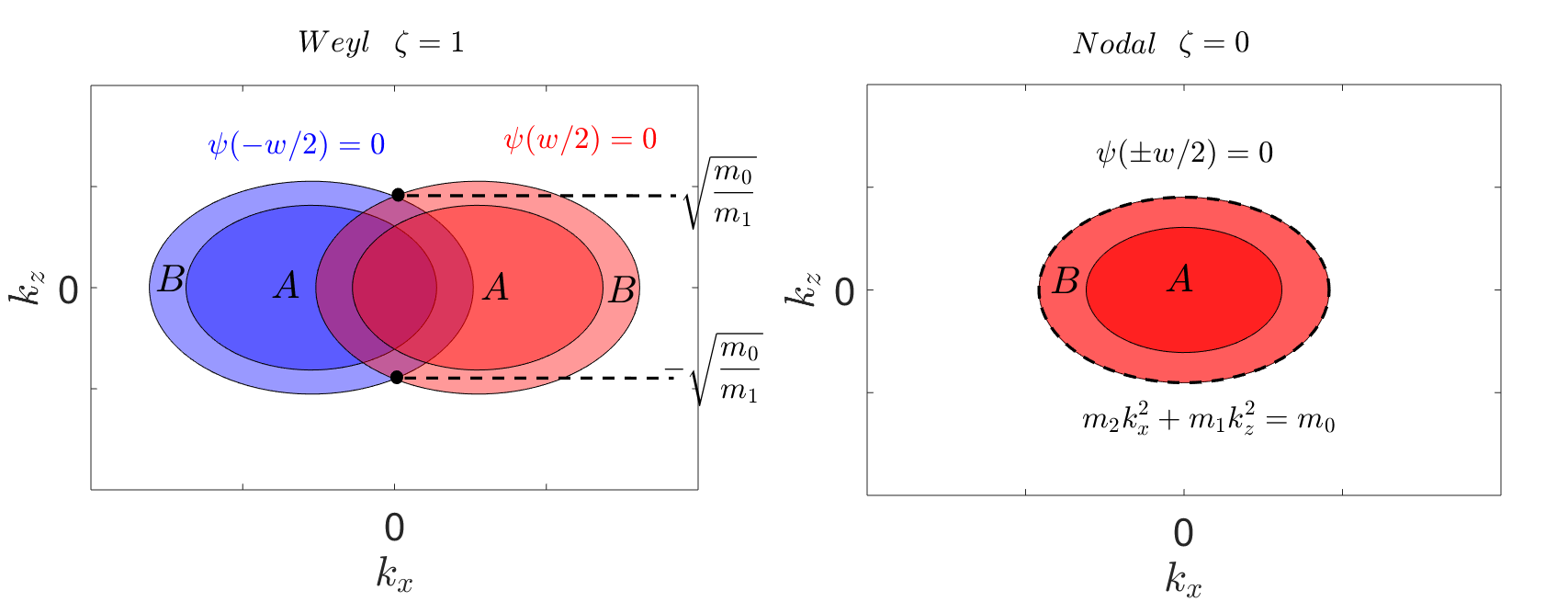}
	\includegraphics[width=0.5\textwidth, height=3cm]{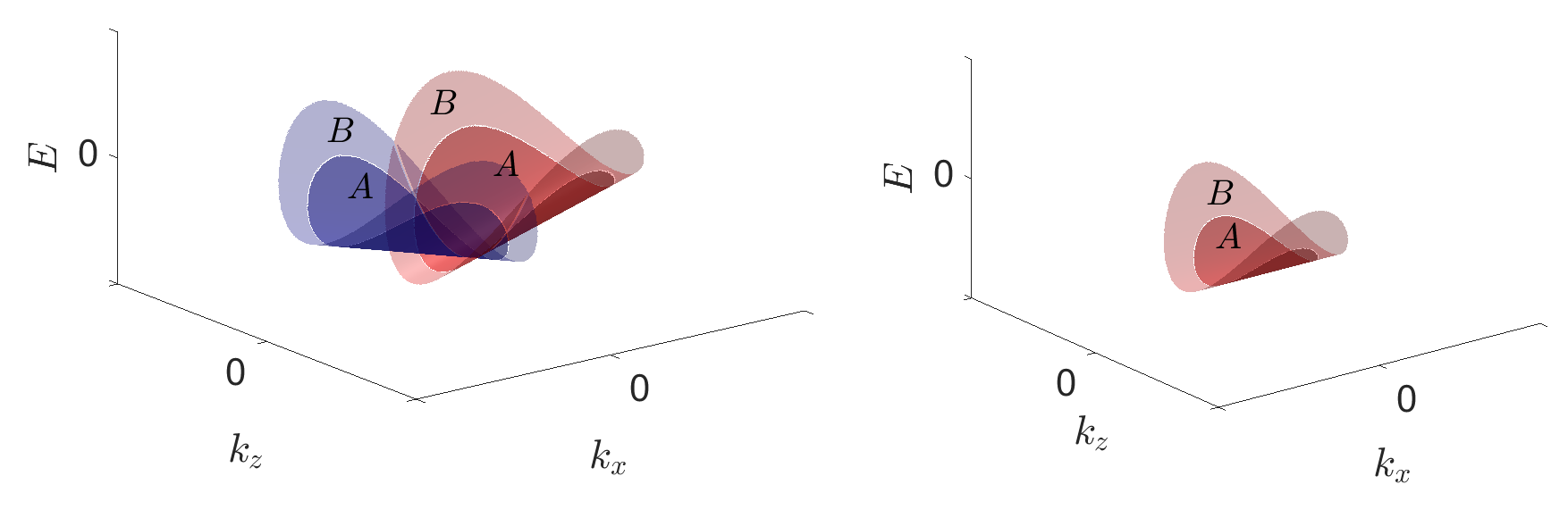}
		\caption{\label{fig:domainofexistenceIEex} A solution to conditions (\ref{trickresultconditionIE2}), (\ref{trickresultconditionIE2c}) for a Weyl semimetal (Left) and a nodal-line semimetal (Right). Regions $B$ correspond to solutions with $\lambda_{1,2}^-\in\mathbb{R}$ (type B Fermi Arcs), while regions $A$ correspond to $\lambda_{1,2}^-\in\mathbb{C}$ (type A Fermi Arcs). Dirac crossings and the line of nodes are also shown respectively for clarity.}
\end{figure}

These two domains represent in general ellipses in the $k_x-k_z$ plane that take a very simple expression in the case of the nodal-line semimetal. The edge of existence is there the elliptical set
$m_2k_x^2+m_1k_z^2=m_0$.
Fig. \ref{fig:domainofexistenceIEex} shows a general domain for Fermi Arcs in the  $k_x-k_z$ plane. ($\zeta=1$)

From (\ref{rootsIE}) the necessary conditions of existence are then:

\begin{equation}
\begin{aligned}
\Delta\in\mathbb{R}\Rightarrow m_2^2>c_2^2
\end{aligned}
\label{existenceconditionIE}
\end{equation}

But the existence of type A Fermi Arcs is much more restrictive. From (\ref{trickresultconditionIE2c}):

\begin{equation}
\begin{aligned}
A\Leftrightarrow\begin{cases}\upsilon<2\sqrt{m_2m_0} &  Weyl\\\upsilon<2\sqrt{m_2m_0-c_2^2\dfrac{m_0}{m_2}} & Nodal\end{cases}
\end{aligned}
\label{existenceconditionIEtypeA}
\end{equation}

\begin{figure}[htb!] 	
	\includegraphics[width=0.5\textwidth]{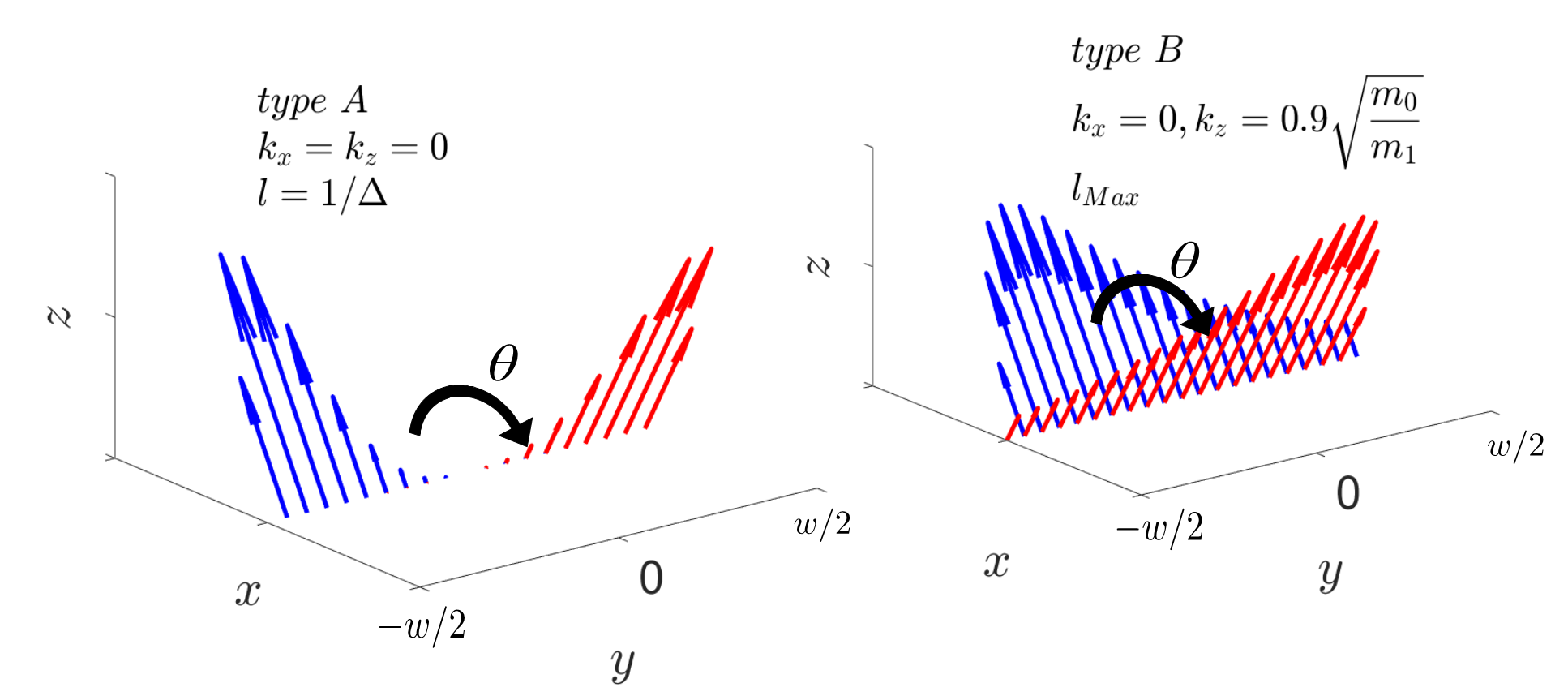}
	\caption{\label{fig:spinorbIENa3Bimodifiedex} Wave functions (\ref{eqn:spinor_IE}) coming from opposite sides in the Isolated edge limit for a Weyl semimetal that hosts type A Fermi Arcs.
	Left: Type A Fermi Arcs. Right: Type B Fermi Arcs.}
\end{figure}

This last condition for the nodal-line semimetal is the complete version of the one found with a simpler model before \cite{Gonzalez2017}.
It is important to emphasize that it is of no importance whether the material is a type I or a type II Weyl semimetal, since the transition to a type II semimetal occurs for $c_1^2>m_1^2$ and these parameters have no influence on the existence conditions (\ref{existenceconditionIE}), (\ref{existenceconditionIEtypeA}).\\
Among other physical quantities of interest, we can obtain the penetration depths and the angle between the spinors:\\
\\
Penetration depth: Type B
\begin{equation}
\begin{aligned}
l_1=\dfrac{1}{\Delta+\sqrt{F}}\Rightarrow\dfrac{1}{2\Delta}\leq l_1\leq\dfrac{1}{\Delta}\\
l_2=\dfrac{1}{\Delta-\sqrt{F}}\Rightarrow\dfrac{1}{\Delta}\leq l_2\leq\infty
\end{aligned}
\label{penetrationIEB}
\end{equation}
Penetration depth: Type A
\begin{equation}
\begin{aligned}
l_1=l_2=\dfrac{1}{\Delta}
\end{aligned}
\label{penetrationIEA}
\end{equation}
Angle between the spinors coming from opposite edges (type A and B)
\begin{equation}
\theta=arcos\left(\dfrac{2c_2^2-m_2^2}{m_2^2}\right)
\label{spinorangleIE}
\end{equation}
Fig.\ref{fig:spinorbIENa3Bimodifiedex} shows the wave functions, including the spinors for a Weyl semimetal and the same parameters as in Fig.\ref{fig:domainofexistenceIEex}.
In the right panel of Fig.\ref{fig:spinorbIENa3Bimodifiedex} the points in $k_x-k_z$ space are very close to the Dirac points and, consequently, to the edge of the existence domains B. Penetration depths can then be quiet large as it is clear from (\ref{penetrationIEB}) so, they can't be properly named "surface states" and, under external perturbations it is expected that some of them are coupled to the bulk. Through (\ref{rootsIE}) and (\ref{penetrationIEB}) it is easy to obtain the domains where Fermi Arcs are close to the surfaces
\begin{equation}
\begin{aligned}
\dfrac{1}{\Delta-\sqrt{F}}\ll w\Rightarrow F\ll \left(\Delta-\dfrac{1}{w}\right)^2\Rightarrow\\ \Rightarrow \dfrac{(k_x\pm \zeta k_{x,0})^2}{R_\zeta^2-2\dfrac{\Delta}{w}+\dfrac{1}{w^2}}+\dfrac{k_z^2}{\left(R^2_\zeta -2\dfrac{\Delta}{w}+\dfrac{1}{w^2}\right)\dfrac{m_2}{m_1}}\ll 1 
\label{eqn:ellipticdomainsSEIE}
\end{aligned}
\end{equation}
This, of course, includes type A regions if they exist.\\
\\
\\

\begin{figure}[htb!]
	\includegraphics[width=0.5\textwidth]{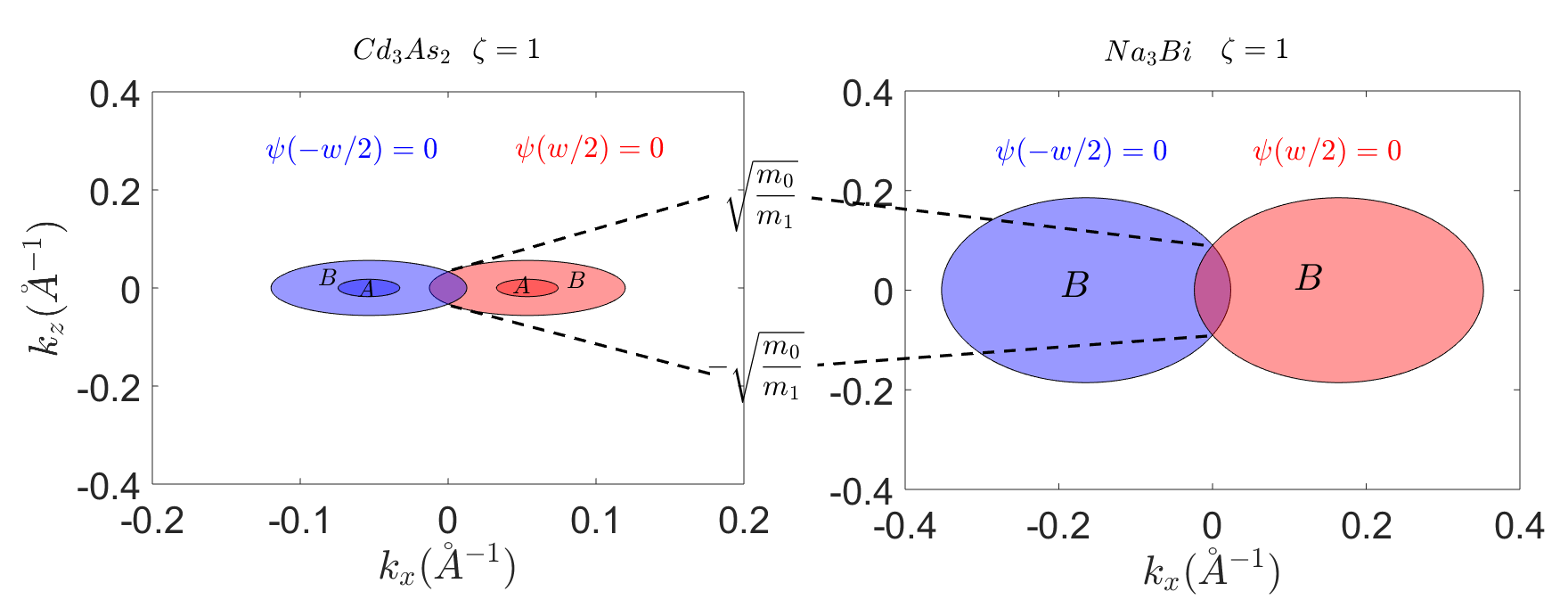}
    \includegraphics[width=0.45\textwidth]{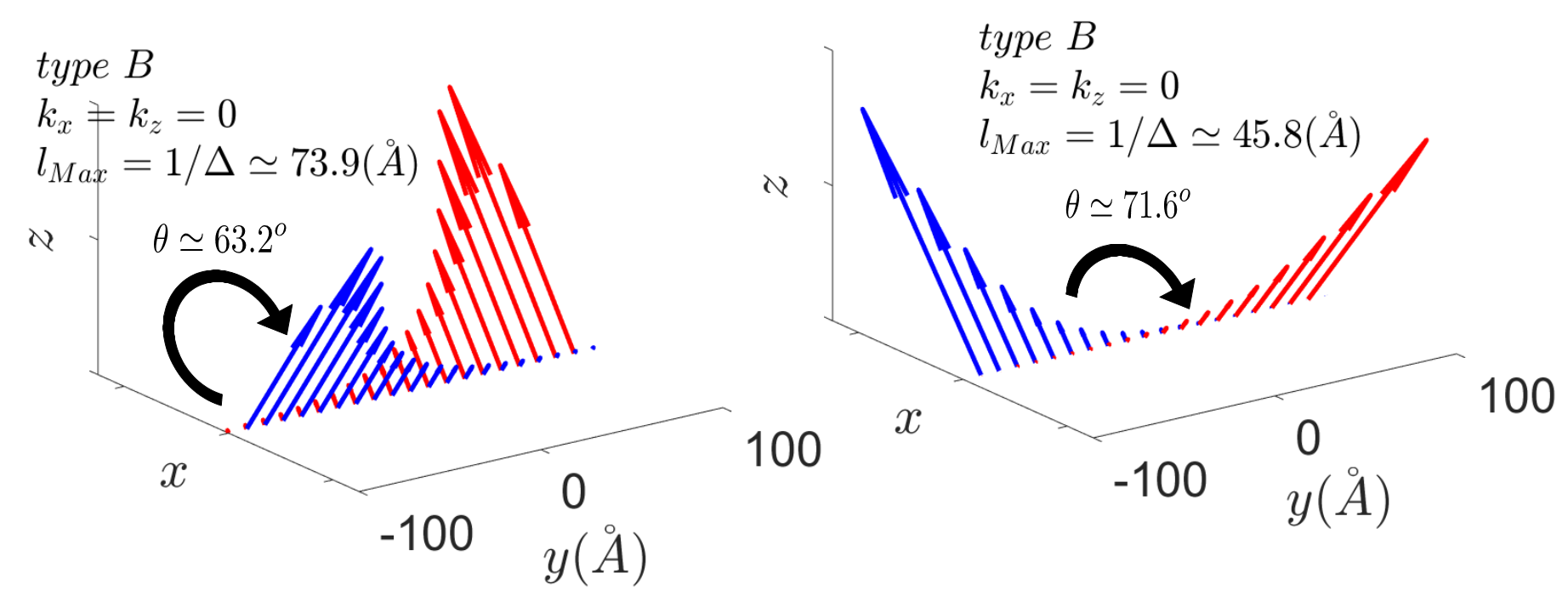}
	\caption{\label{fig:domainofexistenceIEreal} Top panel: Regions of existence for Fermi Arcs in the isolated edge limit for the compounds Cd$_3$As$_2$ (Left) and Na$_3$Bi (Right). Bottom panel: Corresponding spinors, penetration depths, and angles at the domain center $k_x=k_z=0$}
\end{figure}

To obtain real estimates to our results we will use the parameters given by \cite{Wang2012} for A$_3$Bi (A: Na, K, Rb) compounds and by (\textit{cite 2016 D.I. Pkulin}) for Cd$_3$As$_2$. The computed parameters are listed in table \ref{tabla:parametros}.\\
\begin{table}[htb!]
	\begin{center}
		\begin{tabular}{|l|l|}
			\hline
			Na$_3$Bi & Cd$_3$As$_2$ \\
			\hline \hline
			$c_0=-0.06382 \ eV$ & $c_0=-0.0145 \ eV$ \\ \hline
			$c_1=8.7536 \ eV \mathring{A}^2$ & $c_1=10.59 \ eV \mathring{A}^2$ \\ \hline
			$c_2=-8.4008 \ eV \mathring{A}^2$ & $c_2=11.5 \ eV \mathring{A}^2$ \\ \hline
			$m_0=-0.08686 \ eV$ & $m_0=0.0205 \ eV$ \\ \hline
			$m_1=-10.6424 \ eV\mathring{A}^2$ & $m_1=18.77 \ eV\mathring{A}^2$ \\ \hline
			$m_2=-10.3610 \ eV\mathring{A}^2$ & $m_2=13.5 \ eV\mathring{A}^2$ \\ \hline
		\end{tabular}
		\caption{Measured parameters for the compounds Na$_3$Bi and Cd$_3$As$_2$. From \cite{Wang2012} and (\textit{cite 2016 D.I. Pkulin})}
		\label{tabla:parametros}
	\end{center}
\end{table}
The resulting elliptic regions of existence are shown in Fig.\ref{fig:domainofexistenceIEreal}.
Na$_3$Bi does not host type A Fermi Arcs but, Cd$_3$As$_2$ does. The reason is easily understood from (\ref{existenceconditionIEtypeA}): The Fermi velocity $\upsilon$ in Na$_3$Bi is too large or, equivalently, the mass terms $m_2$, $m_0$ are too small. Then, to study properly type A regions in the dispersion relation at fixed $k_x$, we will use a modified model with parameters taken from the Na$_3$Bi compound as a reference, and a decrease in $\upsilon$ of about $40\%$. That would give type A regions, like those shown in Fig. \ref{fig:domainofexistenceIEex}

\begin{figure}[htb!] 	
	\includegraphics[width=0.5\textwidth]{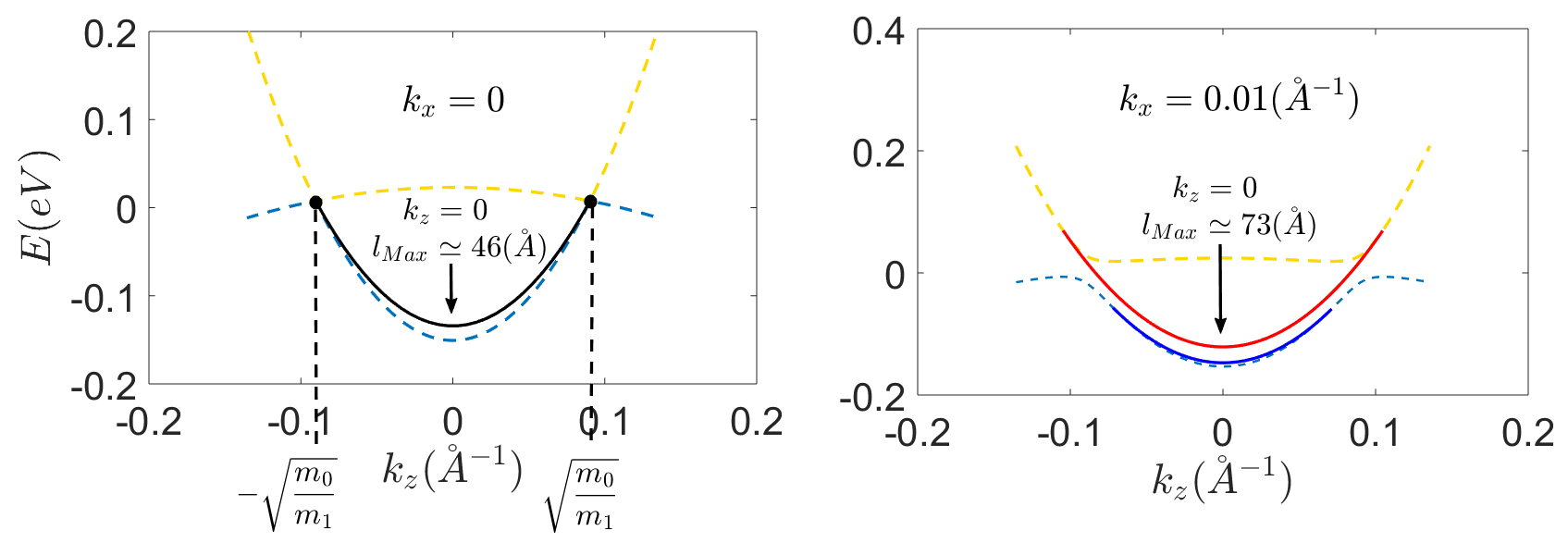}
	\includegraphics[width=0.5\textwidth]{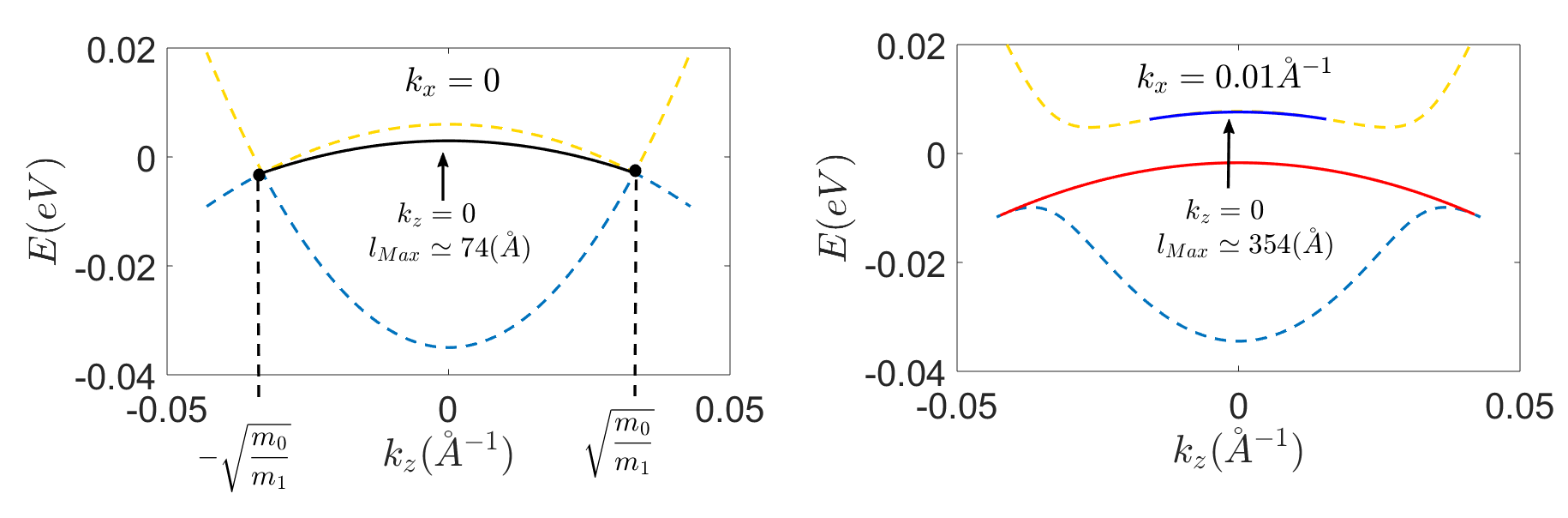}
	\caption{\label{fig:weyldispersion2DIE} Band structure for Weyl semimetals in the isolated edge limit. Dashed lines represents the bulk dispersion relation (\ref{eqn:dispersion}) with the Weyl nodes at $k_z=\pm \sqrt{\dfrac{m_0}{m_1}}$, while the solid lines represents the dispersion (\ref{dispertionIE3}) for Fermi arcs. Top panel: Na$_3$Bi. Bottom panel: Cd$_3$As$_2$. When the dispersion relation for the surface states coming from opposite sides is the same ($k_x=0$, left), the curve is representd as a solid black line. As we move away from $k_x=0$ value, the two curves become shifted by the linear term in $k_x$  (right). The values for $k_x$ are chosen so that two Fermi Arcs exist (intersection of existence domains in Fig.\ref{fig:domainofexistenceIEreal}).}
\end{figure}
 Fig. \ref{fig:weyldispersion2DIE} shows the dispersion (\ref{dispertionIE3}) for the two materials considered and two different values of $k_x$ where the wave functions coming from opposite sides of the slab coexist. 
 
 The JDOS patterns shows particular signatures of these Fermi arcs, depending on the position in $k_x-k_z$ space where these arcs are broken 
 This depends on two main features, namely: 1/ The existence domains in $k_x-k_z$ space and 2/ The existence of type A states. This last property will manifest only in the case of a finite size slab (see \ref{subsec:Finite slab: Type A states}).

 Following Ref. \cite{Kourtis2016}, the JDOS can be defined as
 \begin{equation}
 \begin{aligned}
 J_0=\int d\textbf{k}[A_0(\textbf{k+q}, E)A_0(\textbf{k}, E)]\\
 A_0(\textbf{k}, E)=\dfrac{i}{2\pi}Tr[G(\textbf{k}, E)-G(\textbf{k}, E)^\dag]
 \label{eqn:JDOS}
 \end{aligned}
 \end{equation}
 Where $G(\textbf{k}, E)$ is the Green's function of the system as a function of the energy and momentum. This takes into account just the imaginary part of the Green's function, this is, the density of states (DOS). Hence, the lack of quantum interference effects that depend on the phase of the wave function.

\begin{figure}[htb!]
 	\includegraphics[width=0.5\textwidth]{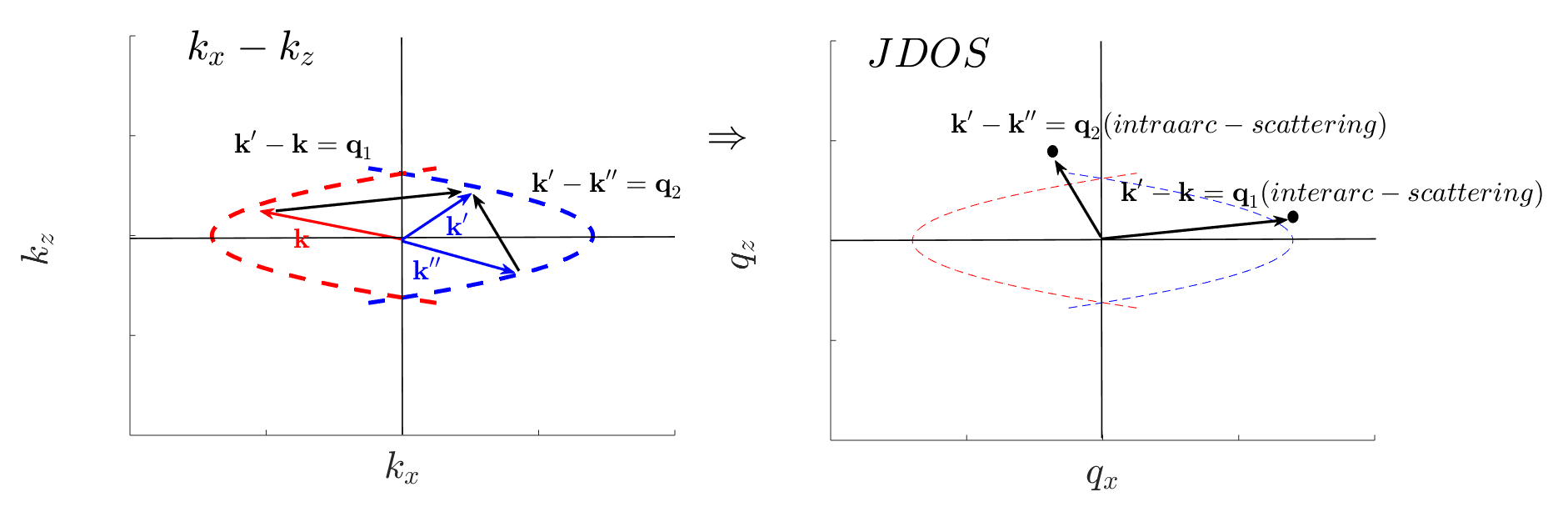}
 	\caption{\label{fig:JDOSex} Computation of the JDOS for intraarc and interarc-scattering}
 \end{figure}

 The JDOS is usually computed numerically with the Green's function of a tight-binding model. However, with the analytic solutions we can use the exact expression (\ref{dispertionIE3}), provided we are inside the existence domains in $k_x-k_z$ space. Then, the JDOS diagrams are just made of points in momentum space, corresponding to vectors that can link two Fermi arcs for the same energy. (Fig. \ref{fig:JDOSex}).

\begin{figure}[htb!]
 	\includegraphics[width=0.5\textwidth]{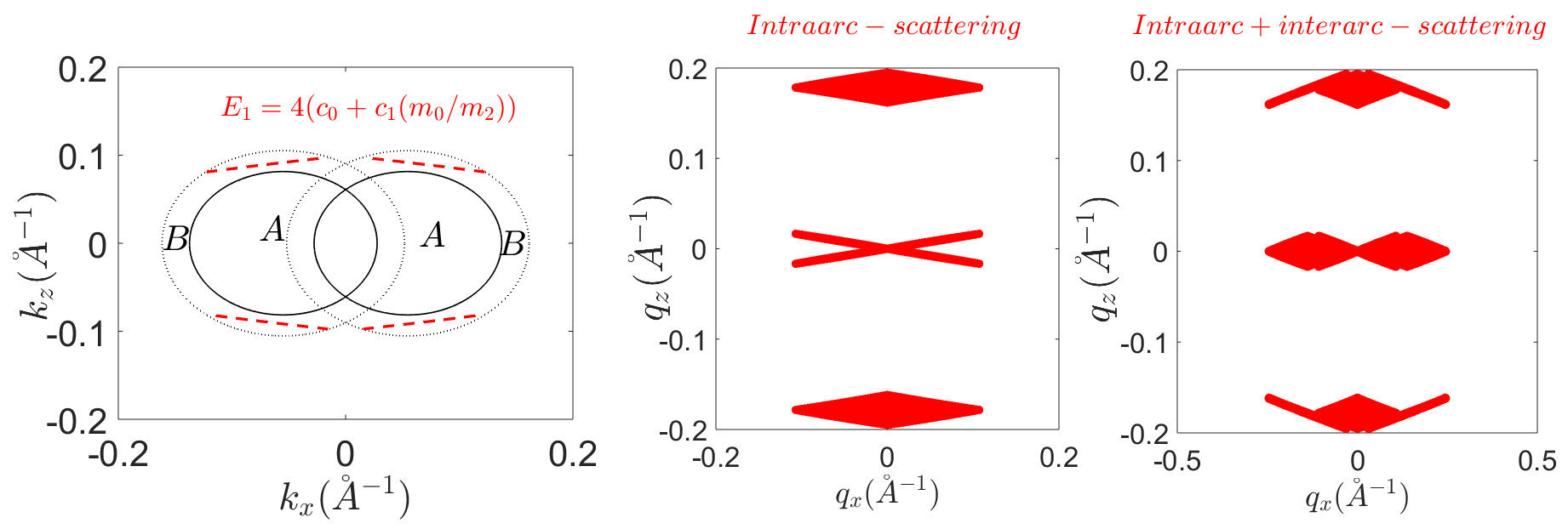}
 	\includegraphics[width=0.5\textwidth]{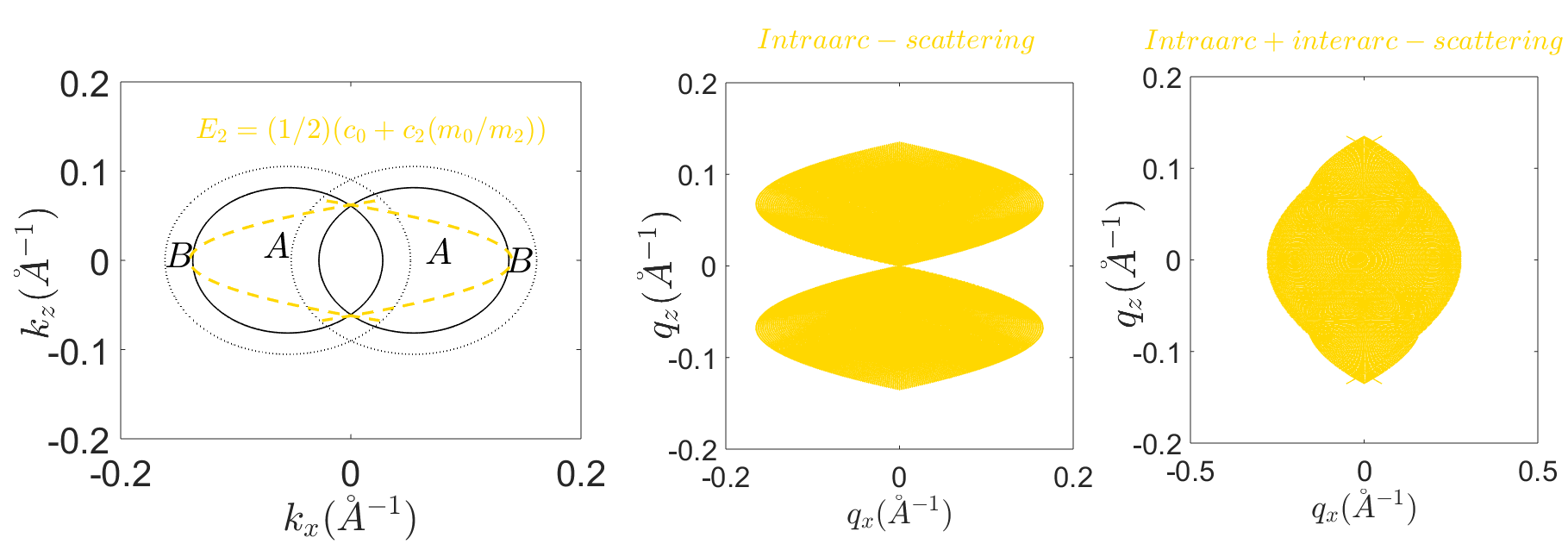}
 	\includegraphics[width=0.5\textwidth]{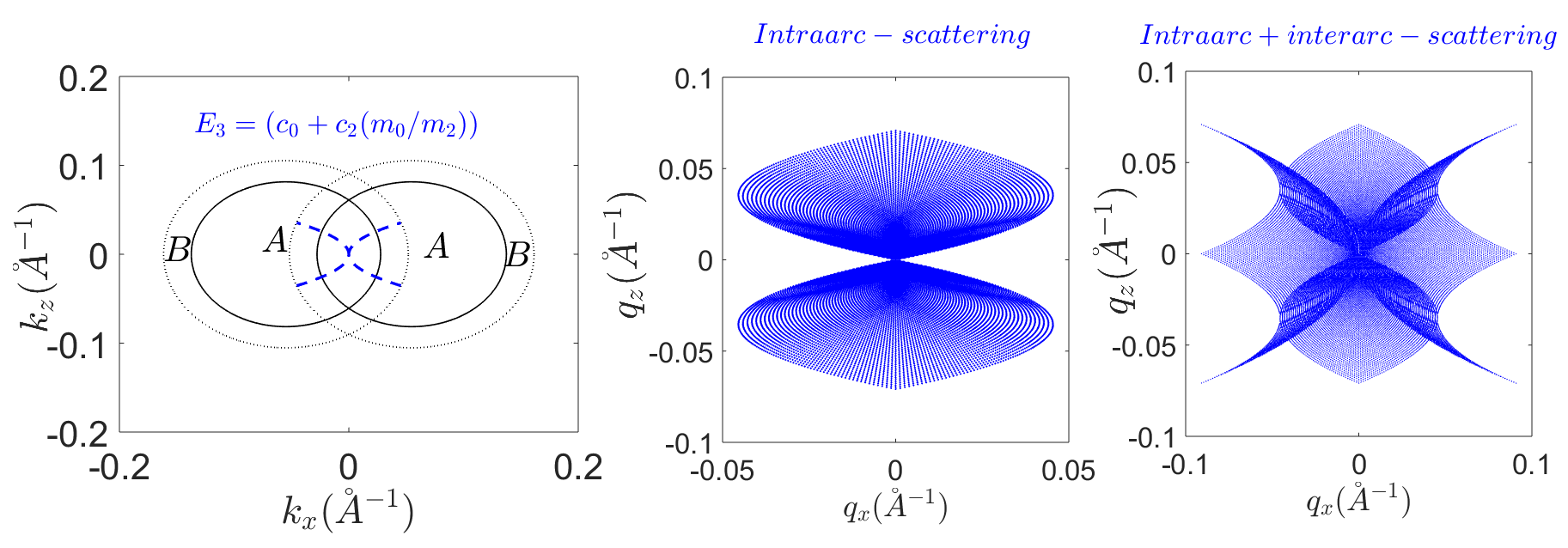}
 	\includegraphics[width=0.5\textwidth]{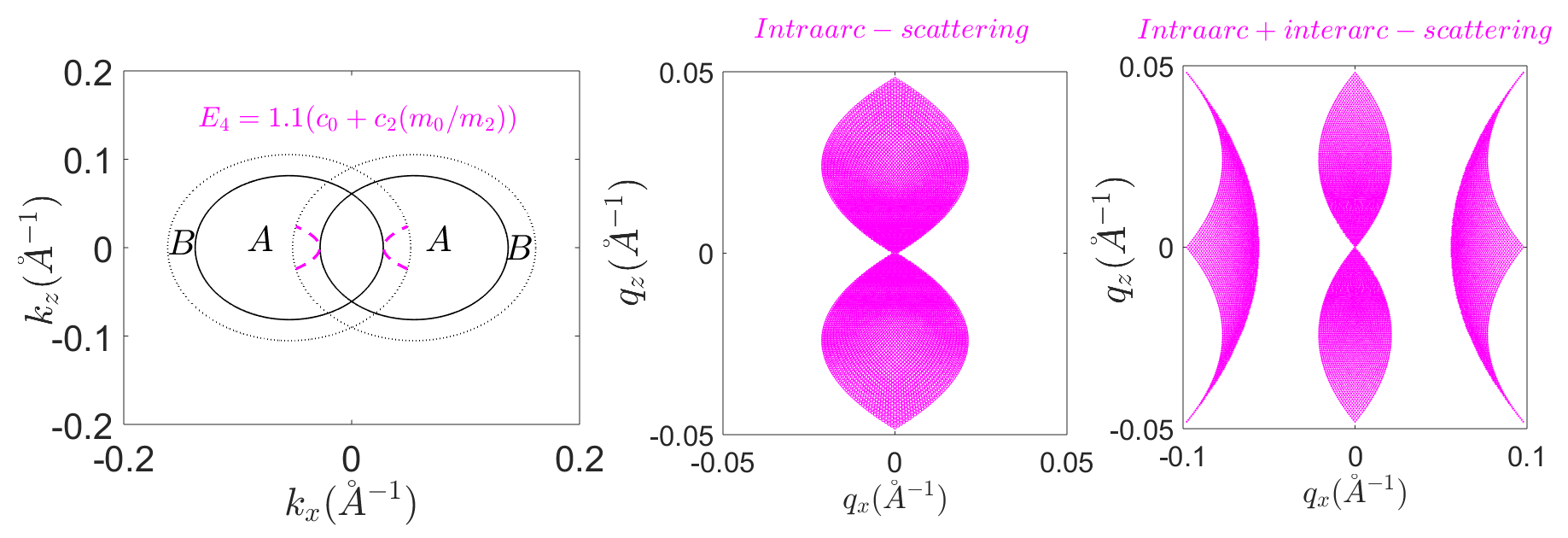}
 \caption{\label{fig:JDOSNa3BimodIE2} Left: Fermi arcs at a given energy in a general domain taking the Na$_3$Bi parameters with a decrease in $\upsilon$ of about $\sim40\%$. The energies are ordered such that $E_1>E_2>E_3>E_4$. Right: Corresponding JDOS diagrams}
 \end{figure}

 Fig.\ref{fig:JDOSNa3BimodIE2} shows these results for different energies in a general situation. In the semi-infinite slab there is no difference, at first, between a model with only type B states and that with type B and type A states. Then the most relevant characteristics not previously stated are the zone boundaries where some of the Fermi arcs are broken. This is specially relevant for the highest energy Fermi arcs (first diagram of Fig.\ref{fig:JDOSNa3BimodIE2}), where the characteristic eight-shaped JDOS for intraarc-scattering disappears, not due to quantum interference but because of the zone boundary.
\subsection{Finite slab: Type B states}
 \label{subsec:Finite slab: Type B states}
\begin{figure}[htp!]
	\includegraphics[width=0.5\textwidth]{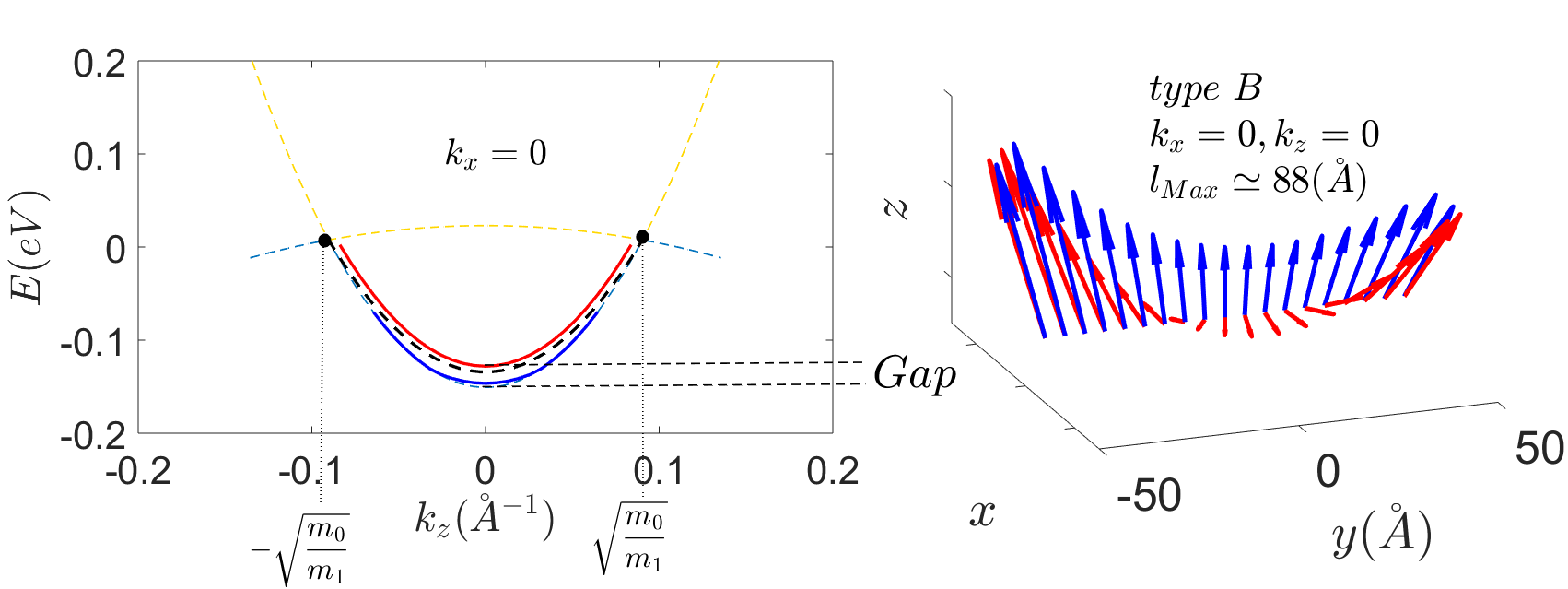}
	\includegraphics[width=0.5\textwidth]{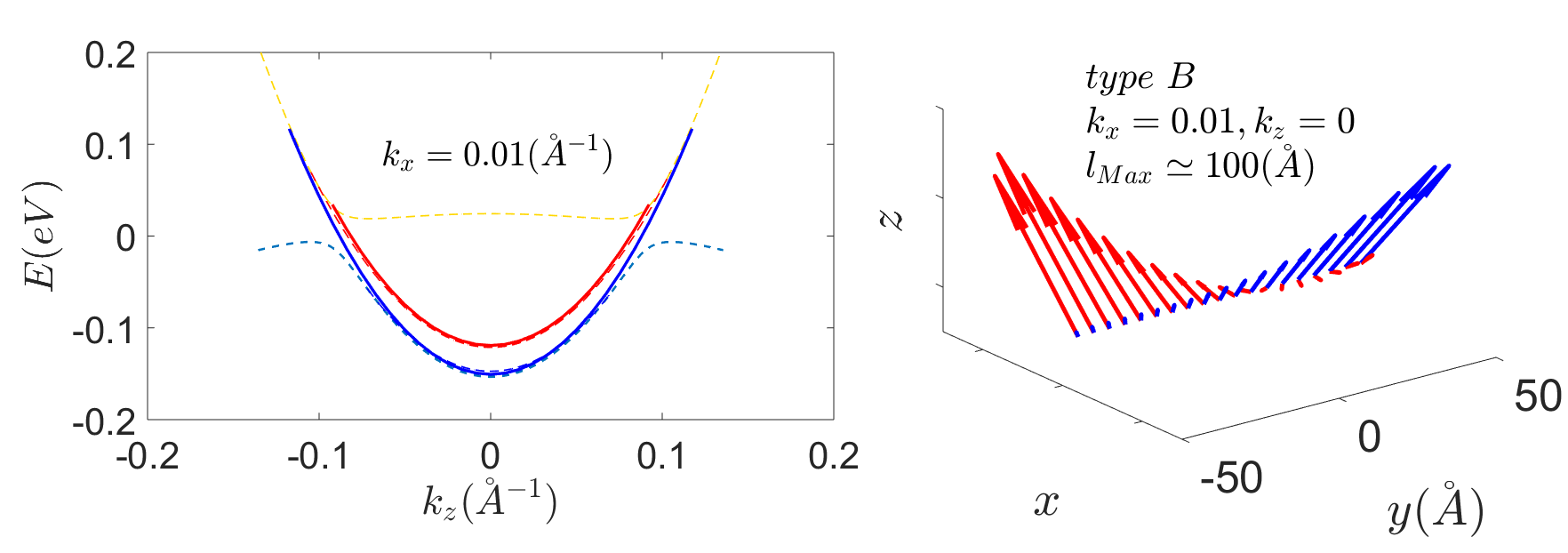}
	\caption{\label{fig:weyldispersion2DslabandspinorNa3Biw100kx0andnum} Finite slab computations 
	for the type B states in the Na$_3$Bi compound. Left: Dispersion as a function of $k_z$ for two different values of $k_x$. $k_x=0$ (Top) and $k_x=0.01 \AA{}$ (Bottom). The solid lines represent the dispersion for the finite slab and dispersions in the isolated edge limit are shown for comparison (dotted lines, red and blue, or black when the two curves coincide). Right: Spinor behavior for $k_z=0$. The size of the slab is w=100$\AA{}$.}
\end{figure}

\begin{figure}[htp!]
	\includegraphics[width=0.5\textwidth]{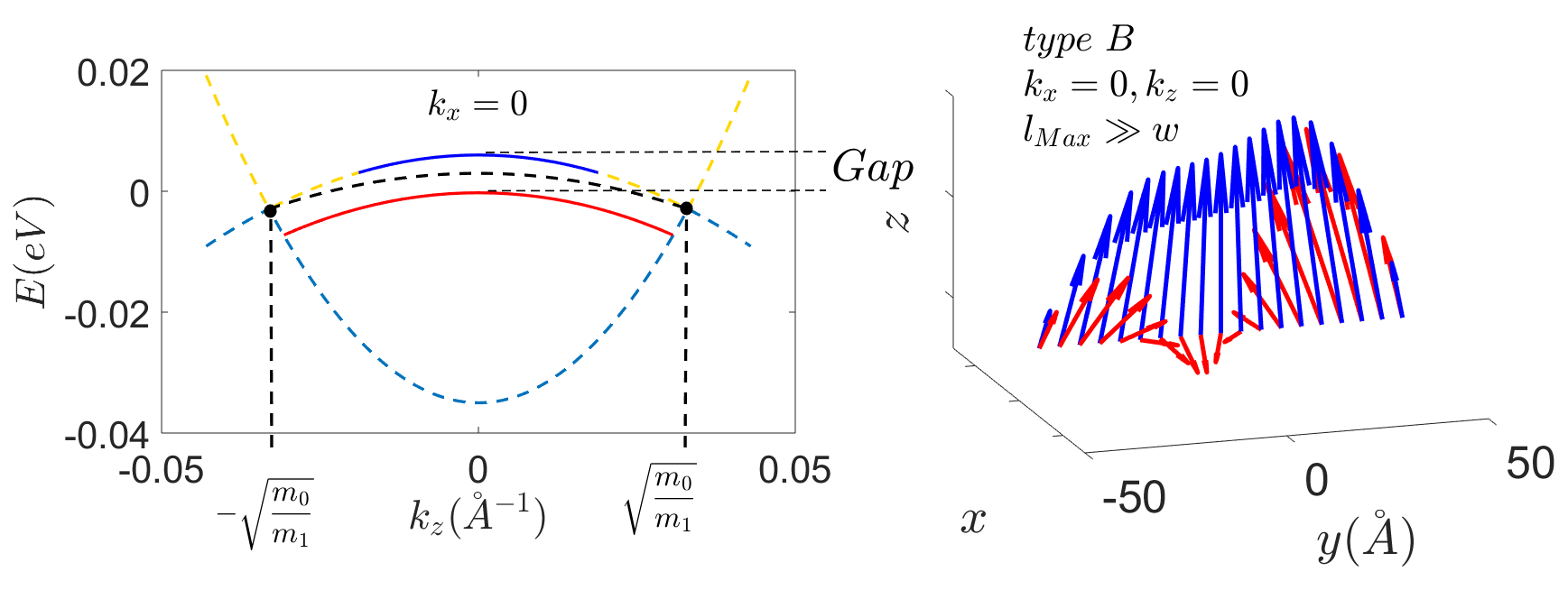}
	\includegraphics[width=0.5\textwidth]{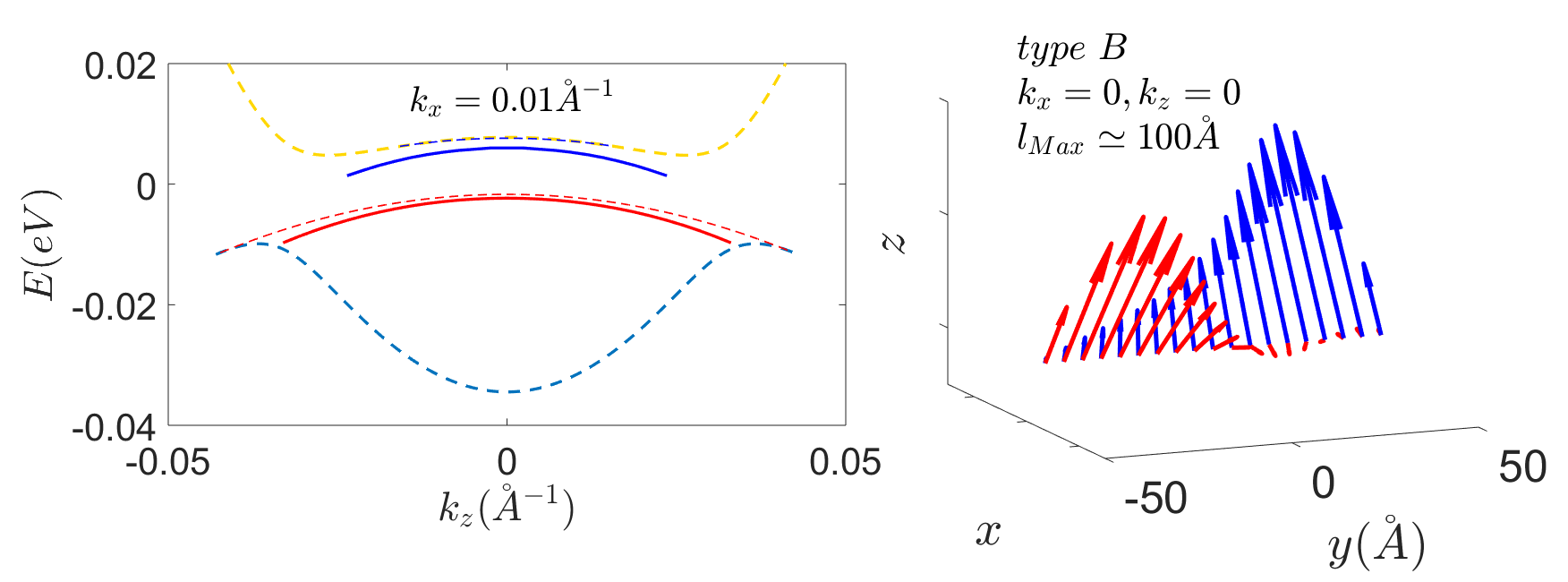}
	\caption{\label{fig:weyldispersion2DslabandspinorCd3As2w1000kx3p2andnum} Finite slab computations
 for the type B states in the Cd$_3$As$_2$ compound. Left: Dispersion as a function of $k_z$ for two different values of $k_x$. $k_x=0$ (Top) and $k_x=0.01 \AA{}$ (Bottom). The solid lines represent the dispersion for the finite slab and dispersions in the isolated edge limit are shown for comparison (dotted lines, red and blue, or black when they coincide). Right: Spinor behavior for $k_z=0$. The size of the slab is w=100$\AA{}$.}
\end{figure}

 For the finite width slab we recover the complete expression  (\ref{eqn:disp3}) together with (\ref{eqn:roots}) and the complete expression for the wave function (\ref{eqn:wavef2}). Then, with a simple fixed point method we can solve the system: 
 \begin{widetext}
 	\begin{equation}
 	\begin{aligned}	
 E=\theta_-+c_{2-}\left[k_x^2-\lambda_1\lambda_2\varGamma\pm\sqrt{\zeta^2k_x^2(\lambda_1^2+\lambda_2^2-2\lambda_1\lambda_2\varGamma)+\lambda_1^2\lambda_2^2(\varGamma^2-1)}\right]\\
 	\lambda_{1,2}=\sqrt{k_y^2-\frac{1}{2c_{2+}c_{2-}}(-b\pm\sqrt{b^2-Q_{W/N}^2})}\\
 	\begin{cases}
 		b=c_{2+}(\theta_- -E)+c_{2-}(\theta_+ -E)-\upsilon^2            \\
 		Q_W^2=4c_{2+}c_{2-}(\theta_- -E)(\theta_+ -E)                   \\
 		Q_N^2=4c_{2+}c_{2-}[(\theta_- -E)(\theta_+ -E)+k_x^2\upsilon^2]
 	\end{cases}\\
 	\label{eqn:nonlinearsystem}
 	\end{aligned}
 	\end{equation}
 \end{widetext}
Then, we can compute exactly the wave function (\ref{eqn:wavef2}):
\begin{widetext}
	\begin{equation*}
	\begin{aligned}
	\psi_{k_x, k_z}(y)_\zeta=A_{1}\left(\Phi_{\lambda_1, \zeta}f(-y)-\dfrac{\sqrt{g^-}\Delta E(-\lambda_1,-\lambda_2)}{\sqrt{g^+}\Delta E(\lambda_1,-\lambda_2)}\Phi_{-\lambda_1, \zeta}f(y)\right)
	=A_{1}\left(\Phi_{-\lambda_1, \zeta}f(y)-\dfrac{\sqrt{g^-}\Delta E(\lambda_1,\lambda_2)}{\sqrt{g^+}\Delta E(-\lambda_1,\lambda_2)}\Phi_{\lambda_1, \zeta}f(-y)\right)
	\end{aligned}
	\end{equation*}
	\\
	With $\Delta E(\lambda_1,\lambda_2)=\theta_--E + c_{2-}[k_x^2+k_x\zeta(\lambda_1 + \lambda_2)+\lambda_1\lambda_2]$ being the difference between the energy in the slab solution and that of the isolated edge solution (see \ref{sec:Fermi_arcs_in_a_slab}, \ref{subsec:isolated_edge} and the appendix)
\end{widetext}
Figs. \ref{fig:weyldispersion2DslabandspinorNa3Biw100kx0andnum} and \ref{fig:weyldispersion2DslabandspinorCd3As2w1000kx3p2andnum} show these results for the two compounds considered, along two lines $k_x=$constant.

The most remarkable result is shown in the dispersion for $k_x=0$ in both compounds. As it is clear from the dispersion curves, the effect of the finite slab is opening a gap for type B states. The size of the energy gap can be analytical estimated for a wide slab as follows:

Taking the expression for the energies (\ref{eqn:disp3}) for $k_x=0$\\
\\ $E(k_x=0)^{\pm}=\theta_-+c_{2-}\left[-\lambda_1\lambda_2\varGamma\pm\lambda_1\lambda_2\sqrt{\varGamma^2-1}\right]\simeq \theta_-+c_{2-}\left[-\lambda_1\lambda_2\varGamma\pm 2\lambda_1\lambda_2\dfrac{\lambda_1+\lambda_2}{\lambda_1-\lambda_2}e^{-\lambda_2w}\right]$\\
\\
\\
In principle, the two solutions depend on different root values but, in the case $k_x=0$ and assuming that their values are not so different than the values for the isolated edge limit, we can take the roots as being the same as in the semi-infinite system (wide slab). This is: $\lambda_1+\lambda_2\to 2\Delta, \lambda_1-\lambda_2\to 2\sqrt{F}, \lambda_1\lambda_2\to \Delta^2-F$.\\
Then:

\begin{equation}
\begin{aligned}
\Delta E_{Gap}=E(k_x=0)^{+}-E(k_x=0)^{-}\simeq\\
\simeq 4c_{2-}(\Delta^2-F)\dfrac{\Delta}{\sqrt{F}}(e^{-\lambda_2^+w}+e^{-\lambda_2^-w})\simeq\\
\simeq 8c_{2-}(\Delta^2-F)\dfrac{\Delta}{\sqrt{F}}e^{-\lambda_2w}
\label{gap}
\end{aligned}
\end{equation}

For the two compounds considered the comparison of this formula with the exact analytic computation gives\\
\\
Na$_3$Bi:\\
\\
$\Delta E_{Gap}=18.4$ meV (Exact) $\Delta E_{Gap}\simeq 16.6$ meV (Formula)\\
\\
Cd$_3$As$_2$:\\
\\
$\Delta E_{Gap}=1.8$ meV (Exact) $\Delta E_{Gap}\simeq 1.7$ meV (Formula)\\
\\
As we will see in next subsection this procedure is not necessary for some of the type A states since they survive exactly with the same structure as in the Isolated edge limit, but in quantized domains of existence.\\


The most important differences in the JDOS diagrams for a finite width slab as compared with the semi-infinite system (Fig. \ref{fig:JDOSNa3BimodIE2}), when there is only type B states, comes from the gap opening.

\begin{figure}[htp!]
	\includegraphics[width=0.5\textwidth]{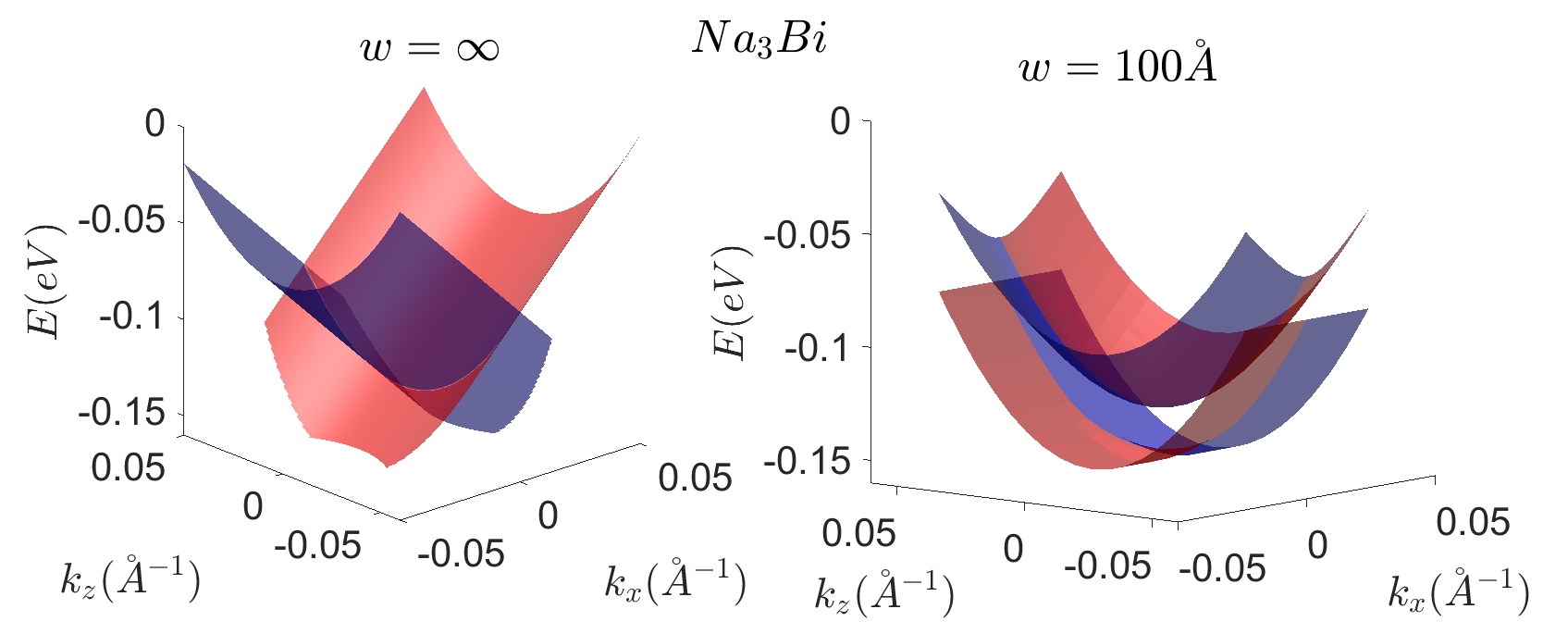}
	\caption{\label{fig:3DdispslabNa3Biw100IEcomp} Na$_3$Bi dispersion surfaces computed from analytic expressions. Left: Isolated edge limit. Right: Finite width slab, $w=100\AA{}$}
\end{figure}

With the analytic expressions (\ref{eqn:nonlinearsystem}) and computing for several $k_x$ values, we can rebuild the dispersion surfaces to understand better these changes.
Fig. \ref{fig:3DdispslabNa3Biw100IEcomp} shows these results compared with the dispersion surfaces in the isolated edge limit for a $100\AA{}$ width slab in the Na$_3$Bi compound.

The gap opening causes the top surface to have a local minimum along the line $k_x=0$ (Fig. \ref{fig:3DdispslabNa3Biw100IEcomp}) so, the approximately parabolic Fermi arcs for a constant energy, don't cross each other in the $k_x-k_z$ plane as in the semi-infinite system (Fig.\ref{fig:JDOSNa3BimodIE2}, yellow diagram).\\
This crossing in the isolated edge limit causes the arcs to touch one another at their vertices for the Energy that corresponds to $k_x=k_z=0$ (Fig. \ref{fig:JDOSNa3BimodIE2}, blue diagram) and, eventually, to separate in momentum space, changing the sign of their original positions in the $k_x$ axis, as the energy decreases (Fig.\ref{fig:JDOSNa3BimodIE2}, pink diagram).

In the JDOS diagrams, this displacement is responsible for the \textit{butterfly-shaped} diagrams in the intraarc+interarc scattering that ends in a JDOS diagram with three different compact domains: The \textit{wings} and the \textit{eight-shaped body} (Fig.\ref{fig:JDOSNa3BimodIE2}, pink diagram).

\begin{figure}[htp!]
	\includegraphics[width=0.5\textwidth]{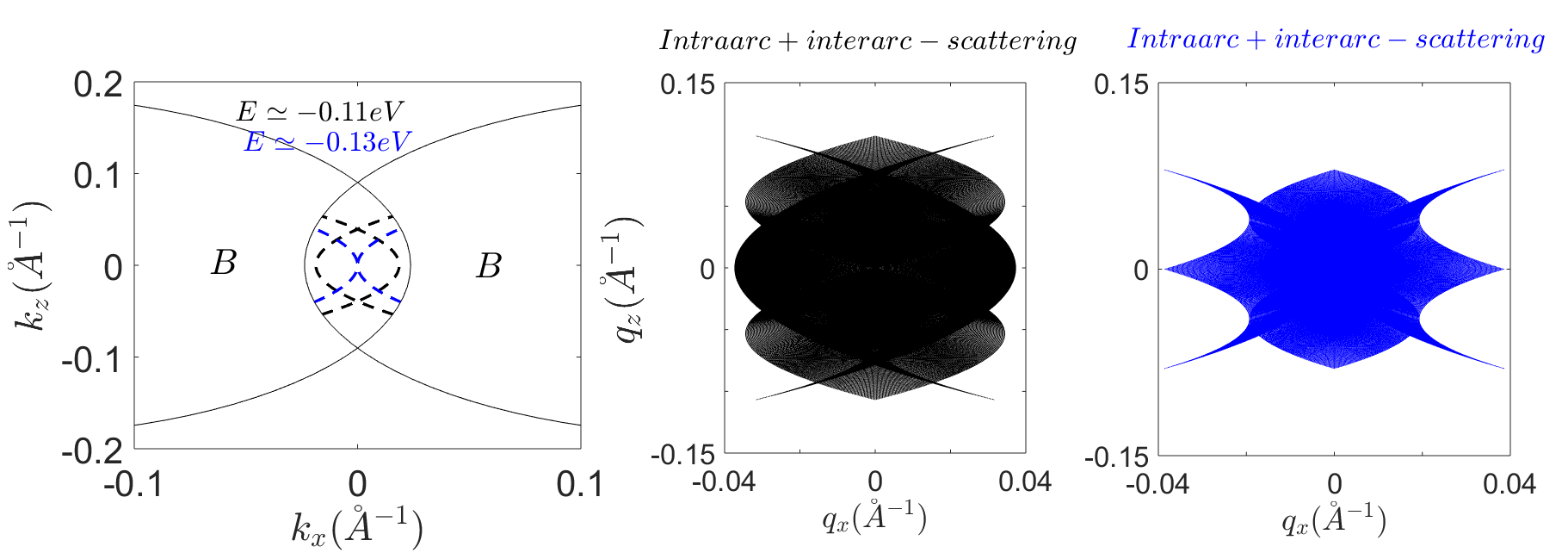}
	\includegraphics[width=0.5\textwidth]{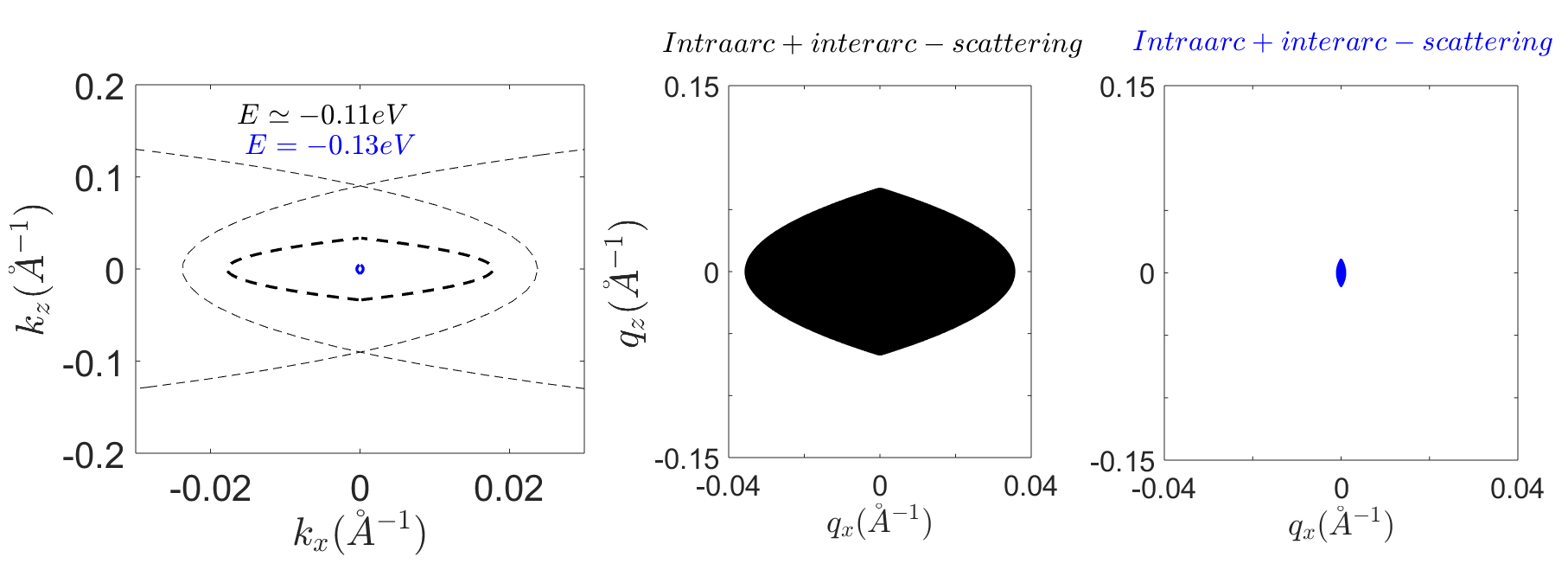}
	\caption{\label{fig:QPINa3Bicomp}Top panel: Fermi arcs and JDOS diagrams when intra-arc and inter-arc scattering is present in the Na$_3$Bi, for two energies close to the $k_x=k_z=0$ point in the isolated edge limit. Bottom panel: Same computations (although the energies are not exactly the same, because of the gap opening), for a slab of width $w=100\AA{}$. The scales in the JDOS diagrams are taken to be equal in these cases just for comparison.}
\end{figure}

Then, for the finite width slab this never happens and the JDOS diagrams, as long as intra-arc and inter-arc scattering are present, have the \textit{diamond-shaped} aspect and finally disappearing when we reach the minimum of the top surface. (Fig.\ref{fig:QPINa3Bicomp}, bottom panel)

 \subsection{Finite slab: Type A states}
 \label{subsec:Finite slab: Type A states}
 When $\lambda_{1,2} \in\mathbb{C}$ the solutions fall in two different categories: Those with a wave function like the type B states (\ref{subsec:Finite slab: Type B states}) with four different exponentials (or two different spinors), that we will call \textquotedblleft type A, four root states\textquotedblright, and those that match the quantization condition $\lambda_2-\lambda_1=\dfrac{2n\pi}{w}i, n\in\mathbb{Z}$  (see appendix). In the later case the solution is exactly that of the isolated edge limit (see \ref{subsec:isolated_edge}) but the Fermi arcs exist in elliptic quantized domains of the $k_x-k_z$ space determined by the condition:
 
 \begin{equation}
 \begin{aligned}
 (k_x\mp \zeta k_{x,0})^2+\dfrac{k_z^2}{\left( \dfrac{m_2}{m_1}\right)}+R_\zeta^2-\Delta^2=\dfrac{-n^2\pi^2}{w^2}\Rightarrow\\
 \Rightarrow\dfrac{(k_x\mp \zeta k_{x,0})^2}{R_\zeta^2-\Delta^2-\dfrac{n^2\pi^2}{w^2}}+\dfrac{k_z^2}{(R_\zeta^2-\Delta^2-\dfrac{n^2\pi^2}{w^2}) \dfrac{m_2}{m_1}}=1 
 \end{aligned}
 \label{eqn:quantizeddomainsslab}
 \end{equation}
 
 We will call these states \textquotedblleft type A, two root states\textquotedblright.\\
 Something really important about these wave functions is that in both cases, \textquotedblleft type A, four root states\textquotedblright or \textquotedblleft type A, two root states\textquotedblright the roots ($\lambda_{1,2}$) or, equivalently, the penetration depths, are exactly the same as in the isolated edge limit (just the real part of the roots, $l=\dfrac{1}{\Delta}$) so, type A and type B states do not mix between them. This is an example of the extreme usefulness of the solution found in subsection \ref{subsec:isolated_edge}, since it is not easy to obtain this result for type A, four root states using the complete formulas (\ref{eqn:nonlinearsystem}).\\
 To sum up, for the type A states we have:
 \\
 \\
\begin{widetext}
 type A, four root states:

 	\begin{equation*}
 	\begin{aligned}
 	\psi_{k_x, k_z}(y)_\zeta=A_{1}\left(\Phi_{\lambda_1, \zeta}f(-y)-\dfrac{\sqrt{g^-}\Delta E(-\lambda_1,-\lambda_2)}{\sqrt{g^+}\Delta E(\lambda_1,-\lambda_2)}\Phi_{-\lambda_1, \zeta}f(y)\right)
 	=\\= A_{1}\left(\Phi_{-\lambda_1, \zeta}f(y)-\dfrac{\sqrt{g^-}\Delta E(\lambda_1,\lambda_2)}{\sqrt{g^+}\Delta E(-\lambda_1,\lambda_2)}\Phi_{\lambda_1, \zeta}f(-y)\right)
 	 \end{aligned}
 	\end{equation*}
 	
 \begin{equation}
 	\begin{aligned}	
 	 f(y)=e^{\lambda_1y}+\dfrac{\text{sinh}(\lambda_1-\lambda_2)w/2}{\text{sinh}(\lambda_2w)}e^{-\lambda_2y}-\dfrac{\text{sinh}(\lambda_1+\lambda_2)w/2}{\text{sinh}(\lambda_2w)}e^{\lambda_2y}\\ \Delta E(\lambda_1,\lambda_2)=\theta_--E + c_{2-}[k_x^2+k_x\zeta(\lambda_1 + \lambda_2)+\lambda_1\lambda_2]\\
 	 \\
 	 \lambda_{1,2}=\Delta \pm iIm(\lambda_{1,2}) \Rightarrow l=\dfrac{1}{\Delta}
 	 \label{eqn:waveftypeAfour}
 	 \end{aligned}
 	\end{equation}
 \end{widetext}
 The energies and the roots given by the expresion (\ref{eqn:nonlinearsystem}). As always, there are two Fermi arcs and consequently, two groups of roots (see end of section \ref{sec: General_solution}) but superscripts $\pm$ are not used in this context since these Fermi arcs match boundary conditions at the two boundaries. In the following case, they obviously match the boundary conditions but they follow the same expressions as the isolated edge states, so we recover the  $\lambda^\pm$ notation.\\
 \begin{widetext}
 	type A, two root states:
 	
 	\begin{equation*}
 	\begin{aligned}
 	\psi_{k_x, k_z}^\pm(y)_\zeta=A_{\mp 1}\begin{pmatrix}\mp 1\\\sqrt{\dfrac{m_2-c_2}{c_2+m_2}}\end{pmatrix}(e^{\lambda_1(\pm y-w/2)}-e^{\lambda_2(\pm y-w/2)}) 
 	\end{aligned}
 	\end{equation*}
 	
 	\begin{equation*}
 	\begin{aligned}
 	\lambda_1^\pm=\Delta+\sqrt{(k_x\mp \zeta k_{x,0})^2+\dfrac{k_z^2}{\left(\dfrac{m_2}{m_1} \right)}+\Delta^2-R_\zeta^2}=\Delta+i\dfrac{n\pi}{w}& & n\in \mathbb{Z}\\
 	\lambda_2^\pm=\Delta-\sqrt{(k_x\mp \zeta k_{x,0})^2+\dfrac{k_z^2}{\left(\dfrac{m_2}{m_1} \right)}+\Delta^2-R_\zeta^2}=\Delta-i\dfrac{n\pi}{w}, & & n\in \mathbb{Z}
 	\end{aligned}
 	\end{equation*}
 	
 	\begin{equation}
 	\begin{aligned}
 	E^\pm=\left(c_0+c_2\left(\dfrac{m_0}{m_2}\right)\right)+\left(c_1-c_2\left(\dfrac{m_1}{m_2}\right)\right)k_z^2\mp\zeta\upsilon\dfrac{\sqrt{m_2^2-c_2^2}}{m_2} k_x\\
 	\Delta=\dfrac{\upsilon}{2\sqrt{m_2^2-c_2^2}}, k_{x,0}=\left(\dfrac{c_2}{m_2}\right)\Delta, R_\zeta^2=\left(\dfrac{m_0}{m_2}\right)+\zeta^2\Delta^2 \left(\dfrac{c_2}{m_2}\right)^2
 	 \label{eqn:waveftypeAtwo}
    \end{aligned}
    \end{equation}
    
 	\end{widetext}
 
 For the type A, two root states, the quantization condition (\ref{eqn:quantizeddomainsslab}) gives a finite number of existence domains for these states, depending on the system's size:
 \begin{equation}
 \begin{aligned}
 R_\zeta^2-\Delta^2-\dfrac{n^2\pi^2}{w^2}>0\Rightarrow n^2<\dfrac{w^2}{pi^2}(R_\zeta^2-\Delta^2)\Rightarrow\\ \Rightarrow n_{Max}=Int\left[\dfrac{w}{\pi}\sqrt{R_\zeta^2-\Delta^2}\right]
 \label{numberofqdomains}
 \end{aligned}
 \end{equation}
 
\begin{figure}[htp!]
 	\includegraphics[width=0.5\textwidth]{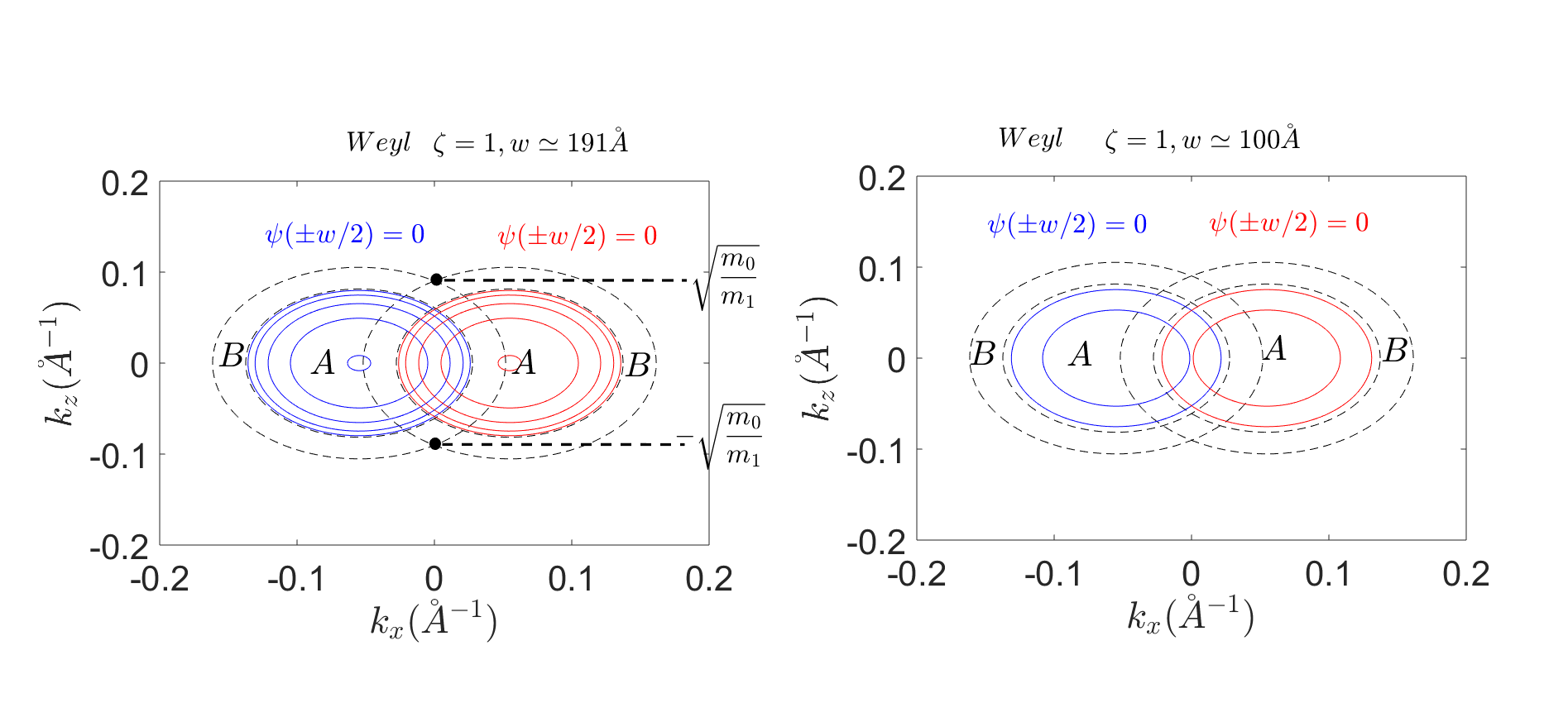}
 	\caption{\label{fig:quantizeddomainsweyl} Domains of existence for type A Fermi arcs in the model with the same parameters as in the Na$_3$Bi compound and a decreased Fermi velocity. Left: w=191$\AA{}$ for this size, expression \ref{eqn:quantizeddomainsslab} generates 5 different elliptic domains of existence for type A, two root states. Right: w= 100$\AA{}$ and there are only two elliptic domains for the two root states.}
 \end{figure}

 To see this quantization and its consequences we, once more, take the model used in \ref{subsec:isolated_edge} to compute the JDOS (Na$_3$Bi parameters with a 40\% decreased Fermi velocity). Fig.\ref{fig:quantizeddomainsweyl} shows these domains for two different sizes.

Now, in the finite slab and always in the case there exist type A, two root states along the line $k_x=0$, type A, four root sates, tend to open an energy gap, just like in the case there are only type B states (see subsec.\ref{subsec:Finite slab: Type B states}). But type A, two root states close this gap at certain points where they exist in the $k_x-k_z$ space, since they follow dispersion (\ref{eqn:waveftypeAtwo}). 

\begin{figure}[htp!]
	\includegraphics[width=0.5\textwidth]{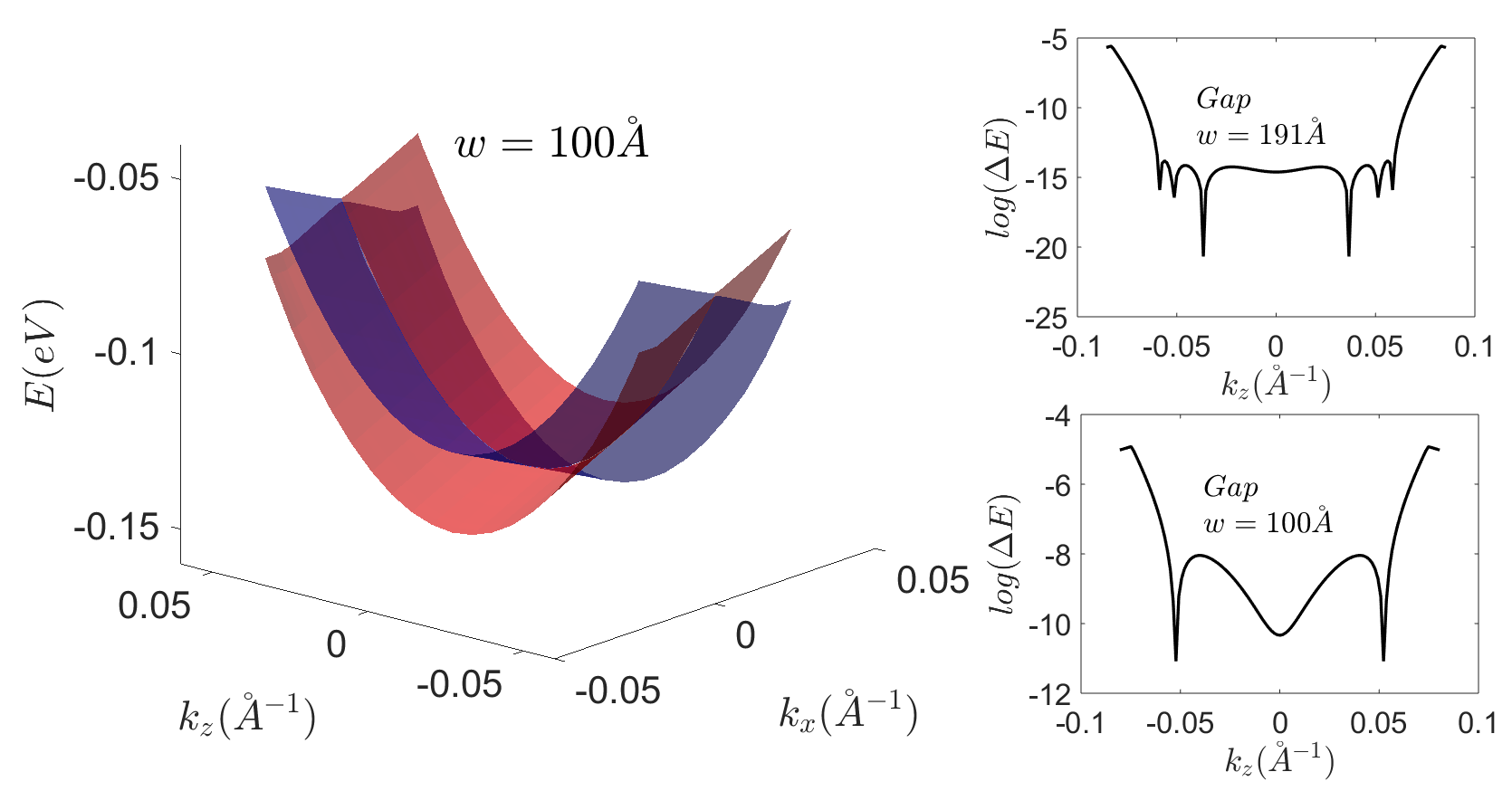}
	\caption{\label{fig:3DdispslabNa3Bimodw100v2}Left: Dispersion surfaces for a narrow slab (w=100$\AA{}$) in the Na$_3$Bi model with a 40\% decreased Fermi velocity. The gap closes at two discrete points where type A, two root states exist (see also Fig.\ref{fig:quantizeddomainsweyl}, right panel) Right: Logarithm of the energy gap for this model and two different sizes w=191$\AA{}$, w=100$\AA{}$.
    Comparison with Fig.\ref{fig:quantizeddomainsweyl} give exact coincidence of these peaks with the regions where the elliptic quantized domains crosses the line $k_x=0$, as expected.}
\end{figure}

This \textit{sewing} effect of the type A, two root states is strong enough to effectively retain the isolated edge limit structure of the Fermi arcs. In Fig. \ref{fig:3DdispslabNa3Bimodw100v2} we show the dispersion surfaces computed analytically in the same way we did with type B states (Fig. \ref{fig:3DdispslabNa3Biw100IEcomp}, Left) for a narrow slab (w=100$\AA$) and our modified Na$_3$Bi model. The gap closing at type A, two root states is strong enough, even in this narrow slab, so that dispersion surfaces (and consequently, Fermi arcs at constant energy) are not affected by finite size effects. The closing at certain points is distinguishable when taking the logarithm of the energy difference. Right panels of Fig.\ref{fig:3DdispslabNa3Bimodw100v2} represents the logarithm of the energy gap for the same two sizes we used in Fig. \ref{fig:quantizeddomainsweyl}. The gap closing takes place where two quantized elliptic domains (\ref{eqn:quantizeddomainsslab}) crosses in $k_x-k_z$ space. (compare with Fig. \ref{fig:quantizeddomainsweyl})

Then, in contrast to a system with only type B states or type A states not along the line $k_x=0$ (as it is the case, respectively, of the two materials considered here)
if a system has type A states along this line, the JDOS is expected to reproduce the diagrams seen in subsec. \ref{subsec:isolated_edge}, Fig. \ref{fig:JDOSNa3BimodIE2}.
Including the \textit{butterfly-shaped} diagrams for low energies (Fig.\ref{fig:JDOSNa3BimodIE2}, blue and pink diagrams) that ends with three different compact domains as the energy decreases.

\section{Conclusions}
\label{sec:conclusions}
In summary, we have computed analytical formulas describing every aspect of the surface states in type I and II Weyl and nodal-line semimetals using a low energy continuous model with slab boundary conditions. 
 The wave functions for the surface states found can be divided in two groups, type A surface states with oscillatory decay into the bulk and type B surface states with simple exponential decay. Type A surface states have particular properties that divides them in two categories: Those that follow the isolated edge limit relations and exist in quantized momentum space domains (type A, two root states) and those that follow the same implicit relations than type B states (type A, four root states). When type A surface states are present along the line $k_x=0$, type A, two root states prevents the gap opening in the energy due to finite size effects, through a closing at discrete points (named \textit{sewing} in the article: Subsection \ref{subsec:Finite slab: Type A states}) where these states exists. This closing continues as the slab width decreases until it is so small that type A, two root states domains disappear. Before we reach this point the \textit{sewing} is strong enough so a qualitative difference in the JDOS for some energies, as compared with the cases where these states don't exist along this line, arises. Then with QPI experiments this qualitative difference could be, in principle, measured. All type A states have the same characteristic decay length $l=\dfrac{1}{\Delta}=\dfrac{2\sqrt{m_2^2-c_2^2}}{\upsilon}$, independently of the material's size so, changing the width of the slab we can control the strength of their coupling with other external systems. Moreover, the number of domains that hold type A, two root states can be changed in the same way. All these physical properties could be important to make them useful for different applications in transport, superlattices and so on.

However, the existence of type A states depends critically on certain model parameters, like the Fermi velocity $\upsilon$. In this sense, the Fermi velocity must be low enough for type A states to appear in the existence domains. Of the two realistic materials studied here, only Cd$_3$As$_2$ match this condition but its Fermi velocity is not low enough to hold type A states within the domain $k_x=0$, where the gap closing gives qualitative differences in the JDOS diagrams.

There are proposals to decrease the effective Fermi velocity in these materials applying electric fields \cite{Diaz-Fernandez2017} and other materials or artificial systems could have different model parameters so they could hold type A states along this domain. For the reasons stated, we think experimental detection of these states in different materials would be very interesting and QPI experiments seem to have the required  resolution and  match the necessary conditions to achieve the detection of type A states.

\acknowledgments
We acknowledge interesting discussions with J. González and finantial support through Spanish grants MINECO/FEDER No. FIS2015-63770-P and No. FIS2014-57432-P and through CAM research consortium QUITEMAD+ S2013/ICE-2801.

\bibliography{Topo_semimetal}

\section{appendix}
\label{sec:appendix}

\subsection{Model, ansatz and general results}
We star with the model Hamiltonian (\ref{eqn:weylhamiltonian}):
\begin{align*}
H=\varepsilon_0(\textbf{k})I+M(\textbf{k})\sigma_z+\hslash\upsilon(\zeta k_x \sigma_x-k_y \sigma_y)
\label{eqn:weylhamiltonianapp}
\end{align*}
\begin{align*}
\varepsilon_0(\textbf{k})=c_0 + c_1k_z^2+c_2(k_x^2+k_y^2)
\end{align*}
\begin{align*}
M(\textbf{k})=m_0-m_1k_z^2-m_2(k_x^2+k_y^2)
\end{align*}

Where $k_z=-i\partial_z$, $k_x=-i\partial_x$, $k_y=-i\partial_y$
Describing a Weyl semimetal ($\zeta=\pm 1$) or a Nodal line semimetal ($\zeta=0$).
In matrix form
\begin{align}
H=\begin{pmatrix}
\varepsilon_0+M & \hslash\upsilon(\zeta k_x+ik_y)\\\hslash\upsilon(\zeta k_x-ik_y) & \varepsilon_0-M 
\end{pmatrix}
\end{align}

Other useful forms:
(In what follows we will fix $\hslash\equiv1$ for simplicity.)
\\
\\
Defining:

\begin{equation}
\begin{cases}c_{2\pm}=c_2\pm m_2\\\theta_{\pm}=(c_0\mp m_0)+(c_1\pm m_1)k_z^2
\end{cases}
\label{eqn:reducedparametersapp}
\end{equation}

Then:

\begin{align}
H=\begin{pmatrix}
c_{2-}(k_x^2+k_y^2)+\theta_{-} & \upsilon(\zeta k_x+ik_y)\\\upsilon(\zeta k_x-ik_y) & c_{2+}(k_x^2+k_y^2)+\theta_{+} 
\end{pmatrix}
\label{eqn:weylhamiltonian3app}
\end{align}

To solve the eigenvalue problem $H\psi=E\psi$ for the bulk states we use a trial wave function $\psi_\zeta(\textbf{r})=e^{i\textbf{k}\cdot\textbf{r}}(\Phi_{\textbf{k},\zeta})$. Then:

\begin{align*}
\begin{vmatrix}
\varepsilon_0+M-E & \upsilon(\zeta k_x+ik_y)\\\upsilon(\zeta k_x-ik_y) & \varepsilon_0-M-E
\end{vmatrix}=0\Rightarrow\\
\Rightarrow E(\textbf{k})=\epsilon_0(\textbf{k})\pm \sqrt{M^2(\textbf{k})+\upsilon^2(\zeta^2k_x^2+k_y^2)}.
\label{eqn:characteristiceqbulkapp}
\end{align*}

The surfaces states must be decaying states from the surface so, for the boundaries along the $k_y$ direction, the trial wave function reads $\psi_\zeta(\textbf{r})=e^{i(k_xx+k_zz)}e^{-\lambda y}(\Phi_{k_x, k_z, \lambda, \zeta})\equiv f_{b}(x, z)\psi_{k_x, k_z}(y)_\zeta$, and we will denote for simplicity: $\psi_{k_x, k_z}(y)_\zeta=e^{-\lambda y}(\Phi_{\lambda, \zeta})$. In the Hamiltonian, this is equivalent to complexification of $k_y$. $k_y\rightarrow i\lambda$. Then in this case:

\begin{widetext}
\begin{equation*}
\begin{aligned}
H=\begin{pmatrix}
\varepsilon_0+M & \upsilon(\zeta k_x-\lambda)\\\upsilon(\zeta k_x+\lambda) & \varepsilon_0-M 
\end{pmatrix}\Rightarrow\begin{vmatrix}
\varepsilon_0+M-E & \upsilon(\zeta k_x-\lambda)\\\upsilon(\zeta k_x+\lambda) & \varepsilon_0-M-E
\end{vmatrix}=0\Rightarrow E=\epsilon_0\pm \sqrt{M^2+\upsilon^2(\zeta^2k_x^2-\lambda^2)}.
\label{eqn:characteristiceqcomplexmomentumapp}
\end{aligned}
\end{equation*}

Notation: $c_z=c_0+c_1k_z^2, m_z=m_0-m_1k_z^2$.\\
Then: $M^2+\upsilon^2(\zeta^2k_x^2-\lambda^2)=(m_z-m_2(k_x^2-\lambda^2))^2+\upsilon^2(\zeta^2k_x^2-\lambda^2))$. And it should be:
$(m_z-m_2(k_x^2-\lambda^2))^2+\upsilon^2(\zeta^2k_x^2-\lambda^2))\geqslant 0$. This is:\\

Weyl semimetal:

\begin{equation*}
\begin{aligned}
\label{eqn:weylfirstpendepthapp}
\lambda_1\geqslant\sqrt{k_x^2+\frac{\upsilon^2}{2m_2^2}-\frac{(m_0-m_1k_z^2)}{m_2}+\sqrt{\left(\frac{\upsilon^2}{2m_2^2}\right)^2-\left(\frac{m_0-m_1k_z^2}{m_2^3}\right)\upsilon^2}}\Rightarrow l_1\leqslant 1/\lambda_1\\
\lambda_2\leqslant\sqrt{k_x^2+\frac{\upsilon^2}{2m_2^2}-\frac{(m_0-m_1k_z^2)}{m_2}-\sqrt{\left(\frac{\upsilon^2}{2m_2^2}\right)^2-\left(\frac{m_0-m_1k_z^2}{m_2^3}\right)\upsilon^2}}\Rightarrow l_2\geqslant 1/\lambda_2
\end{aligned}
\end{equation*}
Nodal line semimetal:
\begin{equation*}
\begin{aligned}
\label{eqn:nodalfirstpendepthapp}
\lambda_1\geqslant\sqrt{k_x^2+\frac{\upsilon^2}{2m_2^2}-\frac{(m_0-m_1k_z^2)}{m_2}+\sqrt{\left(\frac{\upsilon^2}{2m_2^2}\right)^2+\left(\frac{m_2k_x^2-(m_0-m_1k_z^2)}{m_2^3}\right)\upsilon^2}}\Rightarrow l_1\leqslant 1/\lambda_1\\
\lambda_2\leqslant\sqrt{k_x^2+\frac{\upsilon^2}{2m_2^2}-\frac{(m_0-m_1k_z^2)}{m_2}-\sqrt{\left(\frac{\upsilon^2}{2m_2^2}\right)^2+\left(\frac{m_2k_x^2-(m_0-m_1k_z^2)}{m_2^3}\right)\upsilon^2}}\Rightarrow l_2\geqslant 1/\lambda_2
\end{aligned}
\end{equation*}
\end{widetext}
Where $l=1/\lambda$ is the penetration depth in the material.\\
\\
Now, coming back to the Hamiltonian and using parameters given by (\ref{eqn:reducedparametersapp}), we can invert the dispersion relation to compute the roots as a function of the energy, and the components of the spinor $(\Phi_{\lambda, \zeta})$:
\begin{widetext}
\begin{equation*}
\begin{aligned}
(H-E\mathbb{I})(\Phi_{\lambda, \zeta})=0\Rightarrow\begin{pmatrix}
c_{2-}(k_x^2-\lambda^2)+\theta_{-}-E & \upsilon(\zeta k_x-\lambda)\\\upsilon(\zeta k_x+\lambda) & c_{2+}(k_x^2-\lambda^2)+\theta_{+}-E
\end{pmatrix} \begin{pmatrix} \Phi_1\\\Phi_2
\end{pmatrix}_{\lambda, \zeta}=0\\
\\
\left| H-E\mathbb{I}\right|=0\Rightarrow c_{2-}c_{2+}(k_x^2-\lambda^2)+[c_{2-}(\theta_+-E)+c_{2+}(\theta_--E)](k_x^2-\lambda^2)-\upsilon^2(\zeta^2k_x^2-\lambda^2)+(\theta_--E)(\theta_+-E)=0\\
(c_{2-}(k_x^2-\lambda^2)+\theta_{-}-E)\Phi_1+\upsilon(\zeta k_x-\lambda)\Phi_2=0
\label{eqn:weylhamiltonian4app}
\end{aligned}
\end{equation*}
\end{widetext}
So, we have the roots:

\begin{equation}
\label{eqn:rootsapp}
\begin{aligned}
\lambda_{3,4}=-\sqrt{k_x^2-\frac{1}{2c_{2+}c_{2-}}(-b\pm\sqrt{b^2-Q_{W/N}^2})}\\
\lambda_{1,2}=-\lambda_{3,4}
\end{aligned}
\end{equation}

With:

\begin{equation*}
\begin{cases}b=c_{2+}(\theta_- -E)+c_{2-}(\theta_+ -E)-\upsilon^2\\
Q_W^2=4c_{2+}c_{2-}(\theta_- -E)(\theta_+ -E) & Weyl\\
Q_N^2=4c_{2+}c_{2-}[(\theta_- -E)(\theta_+ -E)+k_x^2\upsilon^2] & Nodal\end{cases}
\end{equation*}

And the spinor ($K$ is a normalization constant):

\begin{align}
\Phi_{\lambda, \zeta}=K\begin{pmatrix} \upsilon(\lambda-\zeta k_x)\\c_{2-}(k_x^2-\lambda^2)+(\theta_- -E)\end{pmatrix}=K\begin{pmatrix} \beta_{-\lambda, \zeta}\\ \eta_\lambda \end{pmatrix}
\label{eqn:spinor1app}
\end{align}

In order to have a solution with the proper boundary conditions we need a linear combination with the four possible spinors corresponding to the same energy:

\begin{align}
\psi_{k_x, k_z}(y)_\zeta=\sum_{i=1}^4 A_{i}e^{-\lambda_iy}(\Phi_{\lambda_i, \zeta})
\label{eqn:generalwavef1app}
\end{align}

\subsection{Types of states}
\label{subsec:apptypesofstates}

As we have seen we can write two roots in the following form:

\begin{align*}
\lambda_{1,2}=\sqrt{A\pm B}\\ A=k_x^2+\frac{b}{2c_{2+}c_{2-}} & &   B=\dfrac{1}{2c_{2+}c_{2-}}\sqrt{b^2-Q_{W/N}^2}
\end{align*}

The other two ($\lambda_{3, 4}$) being the same with the opposite sign.
\\
We want to see what are the possible values for $\lambda_{1,2}$. In doing so, we realize that:

\begin{align*}
\lambda_1^2+\lambda_2^2\in\mathbb{R}\\
(\lambda_1^2-\lambda_2^2)^2\in\mathbb{R}
\end{align*}

Then, if we call $\lambda_1=a+ib, \lambda_2=\tilde{a}+i\tilde{b}$ the last condition reads:

\begin{align*}
ab=-\tilde{a}\tilde{b} & & a, \tilde{a}\geqslant0\\
\tilde{a}\tilde{b}[(a^2-b^2)-(\tilde{a}^2-\tilde{b}^2)]=0
\end{align*}

And we have some restrictions over the $\lambda_{1,2}$ values that splits the Hilbert space between surface (Fermi Arcs) and bulk states as a function of the $\lambda_{1,2}$ values. 
\\
\\
Fermi Arcs: (type A) $\lambda_1=a+ib$, $\lambda_2=a-ib$, (type B) $\lambda_1=a$, $\lambda_2=\tilde{a}$. Bulk states: $\lambda_1=ib$, $\lambda_2=i\tilde{b}$. Coupling between bulk and surface: $\lambda_1=a$, $\lambda_2=i\tilde{b}$.

\subsection{Infinite slab: Isolated edges}

In the case of a semi infinite slab we only need a linear combination of two exponentials so the boundary condition $\psi^\pm\xrightarrow[y\to\mp\infty]{} 0$ is automatically satisfied: $\psi_{k_x, k_z}(y)_\zeta^{\pm}=A_{\mp 1}(\Phi_{\mp \lambda_1, \zeta})e^{\pm \lambda_1 y}+A_{\mp 2}(\Phi_{\mp \lambda_2, \zeta})e^{\pm \lambda_2 y}$ where, without loss of generality, $\lambda_{1,2}$ are such that $Re(\lambda_{1,2})\geqslant 0$. Then, the boundary conditions are:
\begin{small}
\begin{equation*}
\begin{aligned}
A_{\mp 1}(\Phi_{\mp \lambda_1, \zeta})e^{\lambda_1 w/2}+A_{\mp 2}(\Phi_{\mp \lambda_2, \zeta})e^{ \lambda_2 w/2}=\begin{pmatrix}0\\0\end{pmatrix}
\end{aligned}
\end{equation*}

This is

\begin{equation*}
\begin{aligned}
\begin{pmatrix} \beta_{\pm \lambda_1, \zeta}e^{\lambda_1 w/2} & \beta_{\pm \lambda_2, \zeta}e^{\lambda_2 w/2}\\\eta_{\lambda_1}e^{\lambda_1 w/2} & \eta_{\lambda_2}e^{\lambda_2 w/2}\end{pmatrix}\begin{pmatrix}A_{\mp 1}\\A_{\mp 2}\end{pmatrix}=\begin{pmatrix}0\\0\end{pmatrix}
\end{aligned}
\end{equation*}

And this is so, if and only if

\begin{equation}
\begin{aligned}
[\beta_{\pm \lambda_1, \zeta}\eta_{\lambda_2}-\beta_{\pm \lambda_2, \zeta}\eta_{\lambda_1}]=\; \; \; \; \; \; \; \; \; \;\\
=\pm \upsilon (\lambda_2-\lambda_1)[\theta_- -E+c_{2-}(k_x^2\pm\zeta k_x(\lambda_1+\lambda_2)+\lambda_1\lambda_2)]=0\\
\text{implying}\; \; \; \; \; \; \; \; \; \; \; \; \; \; \; \; \; \; \; \; \; \; \; \; \; \; \; \; \; \; \; \; \; \; \; \; \; \; \; \; \; \; \; \; \; \; \; \; \; \; \; \; \; \; \; \; \; \; \; \; \; \; \; \; \; \; \; \; \; \; \; \;\\
E^{\pm}=\theta_-+c_{2-}[k_x^2\pm\zeta k_x(\lambda_1+\lambda_2)+\lambda_1\lambda_2]\; \; \; \; \; \; \; \; \;
\label{eqn:dispertionIE1app}
\end{aligned}
\end{equation}
\end{small}
\\
And since the wave function has the normalization constant, solving the system we can write it in the following form

\begin{equation}
\begin{aligned}
\psi_{k_x, k_z}^\pm(y)_\zeta=A_{\mp 1}(\Phi_{\mp\lambda_1, \zeta})(e^{\lambda_1(\pm y-w/2)}-e^{\lambda_2(\pm y-w/2)})
\label{eqn:wavefunctionIE1app}
\end{aligned}
\end{equation}
Now, with dispersion relation (\ref{eqn:dispertionIE1app}) and the expression (\ref{eqn:rootsapp}) for the roots, it is enough to determine explicitly the energy, the roots and the spinor as a function of the model parameters. Instead of this, we will follow and easier way to obtain the desired expressions.\\
Coming back to the Schr\"{o}dinger equation, it is obviously $H\psi_{k_x, k_z}^\pm(y)_\zeta=E\psi_{k_x, k_z}^\pm(y)_\zeta$. And in particular, since the wave function match the boundary conditions  $\left. H\psi_{k_x, k_z}^\pm(y)_\zeta\right|_{\pm w/2}=0$. This give us the following $2\times 2$ system ($k_y=-i\partial_y$)

\begin{small}
\begin{equation*}
\begin{aligned}
\left. \begin{pmatrix}
c_{2-}(k_x^2-\partial_y^2)+\theta_{-} & \upsilon(\zeta k_x+\partial_y)\\\upsilon(\zeta k_x-\partial_y) & c_{2+}(k_x^2-\partial_y^2)+\theta_{+} 
\end{pmatrix}\psi_{k_x, k_z}^\pm(y)_\zeta\right|_{\pm w/2}=\begin{pmatrix}0\\0\end{pmatrix}
\end{aligned}
\end{equation*}

Then:

\begin{equation*}
\begin{aligned}
\begin{pmatrix}c_{2-}(\lambda_2^2-\lambda_1^2) & \pm\upsilon(\lambda_1-\lambda_2)\\\mp\upsilon(\lambda_1-\lambda_2) & c_{2+}(\lambda_2^2-\lambda_1^2)
\end{pmatrix}A_{\mp}(\Phi_{\mp\lambda_1, \zeta})=\begin{pmatrix}0\\0\end{pmatrix}
\end{aligned}
\end{equation*}
Simplifying:
\begin{equation}
\begin{aligned}
\begin{pmatrix}
-c_{2-}(\lambda_1+\lambda_2) & \pm\upsilon\\\mp\upsilon & -c_{2+}(\lambda_1+\lambda_2)
\end{pmatrix}A_{\mp}(\Phi_{\mp\lambda_1, \zeta})=\begin{pmatrix}0\\0\end{pmatrix}
\label{eqn:trick2systemapp}
\end{aligned}
\end{equation}
\end{small}

Again, for a non trivial solution the determinant must be zero so, we have:
\begin{equation}
\begin{aligned}
\lambda_1+\lambda_2=\dfrac{\upsilon}{\sqrt{-c_{2-}c_{2+}}}, &  & A_{\mp}(\Phi_{\mp\lambda_1, \zeta})=A\begin{pmatrix}\mp 1\\ \sqrt{\dfrac{m_2-c_2}{m_2+c_2}}\end{pmatrix}
\label{eqn:tricksol12sysapp}
\end{aligned}
\end{equation}

Next we call $\lambda_{1,2}=\Delta\pm\sqrt{F}$ with $\Delta=\dfrac{\upsilon}{2\sqrt{-c_{2-}c_{2+}}}$. Then, using (\ref{eqn:rootsapp}) we have $\lambda_1^2+\lambda_2^2=2k_x^2+\dfrac{b}{c_{2+}c_{2-}}=\\
=(\lambda_1^2+\lambda_2^2)-2\lambda_1\lambda_2=4\Delta^2-2\lambda_1\lambda_2$, so it is $\lambda_1\lambda_2=\\
=2\Delta^2-k_x^2-\dfrac{b}{2c_{2+}c_{2-}}=-k_x^2-\dfrac{1}{2c_{2+}c_{2-}}(b+\upsilon^2)$. Since it is $b=c_{2+}(\theta_{-}-E)+c_{2-}(\theta_{+}-E)-\upsilon^2=(c_{2+}\theta_{-}+c_{2-}\theta_{+})-(c_{2+}+c_{2-})E-\upsilon^2$ we have for the dispersion relations (\ref{eqn:dispertionIE1app}):\\
\\
$E^{\pm}=\theta_-+c_{2-}[k_x^2\pm\zeta k_x(\lambda_1+\lambda_2)+\lambda_1\lambda_2]=\\
=\theta_{-}+c_{2-}[k_x^2\pm\zeta k_x2\Delta-k_x^2-\dfrac{1}{2c_{2+}c_{2-}}(b+\upsilon^2)]=\\
\\=\theta_{-}+c_{2-}[\pm\zeta k_x2\Delta-\dfrac{1}{2c_{2+}c_{2-}}(b+\upsilon^2)]=\\
=\theta_{-}+c_{2-}[\pm\zeta k_x2\Delta-\dfrac{1}{2c_{2+}c_{2-}}(c_{2+}\theta_{-}+c_{2-}\theta_{+}-(c_{2-}+\\
+c_{2+})E^{\pm})]$ Then: $2c_{2+}E^{\pm}=2c_{2+}\theta_{-}\pm 4c_{2+}c_{2-}\Delta\zeta k_x-\\
\\
-(c_{2+}\theta_{-}+c_{2-}\theta_{+})+(c_{2-}+c_{2+})E^{\pm}$\\
\\
So, the energies

\begin{equation}
\begin{aligned}
E^{\pm}=\dfrac{c_{2+}\theta_{-}-c_{2-}\theta_{+}}{c_{2+}-c_{2-}}\pm \dfrac{4c_{2+}c_{2-}\Delta\zeta k_x}{c_{2+}-c_{2-}}
\label{eqn:dispertionIE2app}
\end{aligned}
\end{equation}

Recovering the original parameters through (\ref{eqn:reducedparametersapp}), we have finally for the energies:

\begin{footnotesize}
\begin{equation}
\begin{aligned}
E^{\pm}=\left(c_0+c_2\left(\dfrac{m_0}{m_2}\right)\right)+\left(c_1-c_2\left(\dfrac{m_1}{m_2}\right)\right)k_z^2\mp\zeta\upsilon\dfrac{\sqrt{m_2^2-c_2^2}}{m_2} k_x
\label{eqn:dispertionIEfapp}
\end{aligned}
\end{equation}
\end{footnotesize}

Next we move onto the expression for the two roots $\lambda_{1,2}$. Since we called $\lambda_{1,2}=\Delta\pm\sqrt{F}$, only $F$ remains to be determined. For this purpose, we back to the expression $\lambda_1\lambda_2=-k_x^2-\dfrac{1}{2c_{2+}c_{2-}}(b+\upsilon^2)=\Delta^2-F=-\dfrac{\upsilon^2}{4c_{2-}c_{2+}}-F$\\
S o: $F=k_x^2+\dfrac{1}{2c_{2+}c_{2-}}(b+\upsilon^2/2)$. But $b=(c_{2+}\theta_{-}+c_{2-}\theta_{+})-(c_{2+}+c_{2-})E^{\pm}-\upsilon^2$ and we have the explicit relation (\ref{eqn:dispertionIE2app}) for the energy, so
\begin{small}
$b=(c_{2+}\theta_{-}++c_{2-}\theta_{+})-(c_{2+}+c_{2-})\left\lbrace \dfrac{c_{2+}\theta_{-}-c_{2-}\theta_{+}}{c_{2+}-c_{2-}}\pm \dfrac{4c_{2+}c_{2-}\Delta\zeta k_x}{c_{2+}-c_{2-}} \right\rbrace--\upsilon^2=\dfrac{2c_{2+}c_{2-}(\theta_{+}-\theta_{-})}{c_{2+}-c_{2-}}\mp \dfrac{4c_{2+}c_{2-}(c_{2+}+c_{2-})\Delta\zeta k_x}{c_{2+}-c_{2-}}-\upsilon^2$ implying: $F=k_x^2+\dfrac{\theta_{+}-\theta_{-}}{c_{2+}-c_{2-}}-\dfrac{\upsilon^2}{4c_{2+}c_{2-}}\mp \dfrac{2(c_{2+}+c_{2-})}{c_{2+}-c_{2-}}\Delta\zeta k_x$. Then: $F=k_x^2+\dfrac{m_1k_z^2-m_0}{m_2}+\Delta^2\mp 2\dfrac{c_2}{m_2}\Delta\zeta k_x=(k_x\mp\mp \dfrac{c_2}{m_2}\zeta\Delta)^2+\dfrac{k_z^2}{\left( \dfrac{m_2}{m_1}\right) }-\left( \dfrac{m_0}{m_2}\right)+\Delta^2-\left( \dfrac{c_2}{m_2}\zeta\Delta\right)^2$
\end{small}

Then, we have $F$ explicitly:

\begin{equation}
\begin{aligned}
F=(k_x\mp \zeta k_{x,0})^2+\dfrac{k_z^2}{\left( \dfrac{m_2}{m_1}\right) }+\Delta^2-R_{\zeta}^2
\label{eqn:Fapp}
\end{aligned}
\end{equation}

Where $R_{\zeta}^2=\left( \dfrac{m_0}{m_2}\right)+\left( \dfrac{c_2}{m_2}\zeta\Delta\right)^2$ and $k_{x,0}=\dfrac{c_2}{m_2}\Delta$\\
Then we have, finally, the desired expression for the two roots (inverse penetration depths)

\begin{equation}
\begin{aligned}
\lambda_1^{\pm}=\Delta+\sqrt{(k_x\mp \zeta k_{x,0})^2+\dfrac{k_z^2}{\left( \dfrac{m_2}{m_1}\right) }+\Delta^2-R_{\zeta}^2}\\
\lambda_1^{\pm}=\Delta-\sqrt{(k_x\mp \zeta k_{x,0})^2+\dfrac{k_z^2}{\left( \dfrac{m_2}{m_1}\right) }+\Delta^2-R_{\zeta}^2}
\label{eqn:rootsIEapp}
\end{aligned}
\end{equation}

So we have two types of states, as stated, depending on the values of $F$\\
\\
\textbf{Type B}
\begin{flushleft}
	$0<F<\Delta^2$
\end{flushleft}
\begin{equation}
\begin{aligned}
\lambda_{1,2}^\pm\in\mathbb{R}\Leftrightarrow
R_\zeta^2-\Delta^2<\dfrac{(k_x\mp \zeta k_{x,0})^2}{R_\zeta^2}+\dfrac{k_z^2}{\left(R_\zeta \sqrt{\dfrac{m_2}{m_1}}\right)^2}<1 
\end{aligned}
\label{trickresultconditionIE2app}
\end{equation}
\textbf{Type A}\\
\begin{flushleft}
	$F<0$ 
\end{flushleft}
\begin{equation}
\begin{aligned}
\lambda_{1,2}^\pm\in\mathbb{C}/  \lambda_1^\pm=(\lambda_2^\pm)^*\Leftrightarrow
\dfrac{(k_x\mp \zeta k_{x,0})^2}{R_\zeta^2-\Delta^2}+\dfrac{k_z^2}{(R_\zeta^2-\Delta^2) \dfrac{m_2}{m_1}}<1 
\end{aligned}
\label{trickresultconditionIE2capp}
\end{equation}

\subsection{Finite slab: Type A states, two root states}
From the above results it is clear that some of the states, in the Type A regions in $k_x-k_z$ space, survive in the slab but in quantized domains.\\
\\
The isolated edge wave function have the structure (\ref{eqn:wavefunctionIE1app}) $\psi_{k_x, k_z}^\pm(y)_\zeta=A_{\mp 1}(\Phi_{\mp\lambda_1, \zeta})(e^{\lambda_1(\pm y-w/2)}-e^{\lambda_2(\pm y-w/2)})$. Then, if we impose $\psi_{k_x, k_z}^\pm(\pm w/2)_\zeta=0\Rightarrow(e^{\lambda_1w}-e^{\lambda_2w})=0\Rightarrow \lambda_2-\lambda_1=\dfrac{2n\pi}{w}i, n\in\mathbb{Z}\Rightarrow \sqrt{F}=\dfrac{n\pi}{w}i\Leftrightarrow F=\dfrac{-n^2\pi^2}{w^2}$ Then, this quantized regions are determined by

\begin{equation}
\begin{aligned}
(k_x\mp \zeta k_{x,0})^2+\dfrac{k_z^2}{\left( \dfrac{m_2}{m_1}\right)}+R_\zeta^2-\Delta^2=\dfrac{-n^2\pi^2}{w^2}\Rightarrow\\
\Rightarrow\dfrac{(k_x\mp \zeta k_{x,0})^2}{R_\zeta^2-\Delta^2-\dfrac{n^2\pi^2}{w^2}}+\dfrac{k_z^2}{(R_\zeta^2-\Delta^2-\dfrac{n^2\pi^2}{w^2}) \dfrac{m_2}{m_1}}=1 
\end{aligned}
\label{quantizeddomainsIEapp}
\end{equation}
The number of these quantized regions depends in this way in system's size, since it must be $R_\zeta^2-\Delta^2-\dfrac{n^2\pi^2}{w^2}>0\Rightarrow n^2<\dfrac{w^2}{pi^2}(R_\zeta^2-\Delta^2)\Rightarrow n_{Max}=Int\left[\dfrac{w}{\pi}\sqrt{R_\zeta^2-\Delta^2}\right]$
\subsection{Finite slab: Type B states and type A, four root states}
We will impose boundary conditions in a slab geometry of width $w$ such that $\psi_{k_x, k_z}(-w/2)_\zeta=\psi_{k_x, k_z}(w/2)_\zeta=0$. These conditions are\\
\\
\begin{footnotesize}
$A_{1}\Phi_{\lambda_1, \zeta}e^{-\lambda_1w/2}+A_{2}\Phi_{\lambda_2, \zeta}e^{-\lambda_2w/2}+
+A_{-1}\Phi_{-\lambda_1, \zeta}e^{\lambda_1w/2}
++A_{-2}\Phi_{-\lambda_2, \zeta}e^{\lambda_2w/2}=0$\\
\\
$A_{1}\Phi_{\lambda_1, \zeta}e^{\lambda_1w/2}+A_{2}\Phi_{\lambda_2, \zeta}e^{\lambda_2w/2}+
+A_{-1}\Phi_{-\lambda_1, \zeta}e^{-\lambda_1w/2}
++A_{-2}\Phi_{-\lambda_2, \zeta}e^{-\lambda_2w/2}=0$
\end{footnotesize}
If we sum and substract these two conditions they can be transformed in a more convenient way\\
\\
\begin{footnotesize}
$\text{cosh}(\lambda_1w/2)[A_{1}\Phi_{\lambda_1, \zeta}+A_{-1}\Phi_{-\lambda_1, \zeta}]+
\text{cosh}(\lambda_2w/2)[A_{2}\Phi_{\lambda_2, \zeta}++A_{-2}\Phi_{-\lambda_2, \zeta}]=0$\\
\\
$\text{sinh}(\lambda_1w/2)[A_{1}\Phi_{\lambda_1, \zeta}-A_{-1}\Phi_{-\lambda_1, \zeta}]+
+\text{sinh}(\lambda_2w/2)[A_{2}\Phi_{\lambda_2, \zeta}--A_{-2}\Phi_{-\lambda_2, \zeta}]=0$
\end{footnotesize}

\begin{widetext}
Multiplying the first expression by $\text{sinh}(\lambda_2w/2)$, the second one by $\text{cosh}(\lambda_2w/2)$, summing and subtracting again, we find

\begin{equation*}
\begin{aligned}
A_{2}\Phi_{\lambda_2, \zeta}=A_{-1}\Phi_{-\lambda_1, \zeta}\dfrac{\text{sinh}((\lambda_1-\lambda_2)w/2)}{\text{sinh}(\lambda_2w)}-A_{1}\Phi_{\lambda_1, \zeta}\dfrac{\text{sinh}((\lambda_1+\lambda_2)w/2)}{\text{sinh}(\lambda_2w)}\\
A_{-2}\Phi_{-\lambda_2, \zeta}=A_{1}\Phi_{\lambda_1, \zeta}\dfrac{\text{sinh}((\lambda_1-\lambda_2)w/2)}{\text{sinh}(\lambda_2w)}-A_{-1}\Phi_{-\lambda_1, \zeta}\dfrac{\text{sinh}((\lambda_1+\lambda_2)w/2)}{\text{sinh}(\lambda_2w)}
\end{aligned}
\end{equation*}

Then we can express the wave function as a function of two spinors $\Phi_{\lambda_1, \zeta}$ and $\Phi_{-\lambda_1, \zeta}$.

The general solution for the wave function (\ref{eqn:generalwavef1app}) reads:
\begin{equation}
\begin{aligned}
\psi_{k_x, k_z}(y)_\zeta=A_{1}\Phi_{\lambda_1, \zeta}\lbrace e^{-\lambda_1y}+\dfrac{1}{\text{sinh}(\lambda_2w)}[\text{sinh}((\lambda_1-\lambda_2)w/2)e^{\lambda_2y}-\text{sinh}((\lambda_1+\lambda_2)w/2)e^{-\lambda_2y}]\rbrace+\\
+A_{-1}\Phi_{-\lambda_1, \zeta}\lbrace e^{\lambda_1y}+\dfrac{1}{\text{sinh}(\lambda_2w)}[\text{sinh}((\lambda_1-\lambda_2)w/2)e^{-\lambda_2y}-\text{sinh}((\lambda_1+\lambda_2)w/2)e^{\lambda_2y}]\rbrace
\end{aligned}
\label{eqn:wavef2app}
\end{equation}
or taking $f(y)=e^{\lambda_1y}+\dfrac{1}{\text{sinh}(\lambda_2w)}[\text{sinh}((\lambda_1-\lambda_2)w/2)e^{-\lambda_2y}-\text{sinh}((\lambda_1+\lambda_2)w/2)e^{\lambda_2y}]$:

\begin{equation}
\begin{aligned}
\psi_{k_x, k_z}(y)_\zeta=A_{1}\Phi_{\lambda_1, \zeta}f(-y)+A_{-1}\Phi_{-\lambda_1, \zeta}f(y)
\end{aligned}
\label{eqn:wavef3}
\end{equation}
\\
The expressions for $\Phi_{\lambda_2, \zeta}$ and $\Phi_{-\lambda_2, \zeta}$ give us the necessary conditions for a non-trivial solution through a 4$\times$4 system of equations:

\begin{equation*}
\begin{aligned}
A_{2}\Phi_{\lambda_2, \zeta}=A_{-1}\Phi_{-\lambda_1, \zeta}\dfrac{\text{sinh}((\lambda_1-\lambda_2)w/2)}{\text{sinh}(\lambda_2w)}-A_{1}\Phi_{\lambda_1, \zeta}\dfrac{\text{sinh}((\lambda_1+\lambda_2)w/2)}{\text{sinh}(\lambda_2w)}\\
A_{-2}\Phi_{-\lambda_2, \zeta}=A_{1}\Phi_{\lambda_1, \zeta}\dfrac{\text{sinh}((\lambda_1-\lambda_2)w/2)}{\text{sinh}(\lambda_2w)}-A_{-1}\Phi_{-\lambda_1, \zeta}\dfrac{\text{sinh}((\lambda_1+\lambda_2)w/2)}{\text{sinh}(\lambda_2w)}
\end{aligned}
\end{equation*}
Using (\ref{eqn:spinor1app}):

\begin{align}
\begin{pmatrix}
\beta_{-\lambda_1,\zeta}\dfrac{\text{sinh}((\lambda_1+\lambda_2)w/2)}{\text{sinh}(\lambda_2w)}&
\beta_{-\lambda_2,\zeta}& -\beta_{\lambda_1,\zeta}\dfrac{\text{sinh}((\lambda_1-\lambda_2)w/2)}{\text{sinh}(\lambda_2w)}&0\\\eta_{\lambda_1}\dfrac{\text{sinh}((\lambda_1+\lambda_2)w/2)}{\text{sinh}(\lambda_2w)} & \eta_{\lambda_2}&-\eta_{\lambda_1}\dfrac{\text{sinh}((\lambda_1-\lambda_2)w/2)}{\text{sinh}(\lambda_2w)}&0\\-\beta_{-\lambda_1,\zeta}\dfrac{\text{sinh}((\lambda_1-\lambda_2)w/2)}{\text{sinh}(\lambda_2w)}&0&\beta_{\lambda_1,\zeta}\dfrac{\text{sinh}((\lambda_1+\lambda_2)w/2)}{\text{sinh}(\lambda_2w)}&\beta_{\lambda_2,\zeta}\\-\eta_{\lambda_1}\dfrac{\text{sinh}((\lambda_1-\lambda_2)w/2)}{\text{sinh}(\lambda_2w)}&0&\eta_{\lambda_1}\dfrac{\text{sinh}((\lambda_1+\lambda_2)w/2)}{\text{sinh}(\lambda_2w)}&\eta_{\lambda_2}
\end{pmatrix}\begin{pmatrix} A_1\\A_2\\A_{-1}\\A_{-2}
\end{pmatrix}=\begin{pmatrix} 0\\0\\0\\0\end{pmatrix}
\label{eqn:fundamentalsystem1}
\end{align}

After some manipulations, this system can be reduced to:

\begin{align*}
\begin{pmatrix}
[\beta_{-\lambda_1,\zeta}\eta_{\lambda_2}-\beta_{-\lambda_2,\zeta}\eta_{\lambda_1}]\text{sinh}((\lambda_1+\lambda_2)w/2)&
0&[\beta_{-\lambda_2,\zeta}\eta_{\lambda_1}-\beta_{\lambda_1,\zeta}\eta_{\lambda_2}]\text{sinh}((\lambda_1-\lambda_2)w/2)&0\\ 
\eta_{\lambda_1}\dfrac{\text{sinh}((\lambda_1+\lambda_2)w/2)}{\text{sinh}(\lambda_2w)} & \eta_{\lambda_2}&-\eta_{\lambda_1}\dfrac{\text{sinh}((\lambda_1-\lambda_2)w/2)}{\text{sinh}(\lambda_2w)}&0\\
[\beta_{\lambda_2,\zeta}\eta_{\lambda_1}-\beta_{-\lambda_1,\zeta}\eta_{\lambda_2}]\text{sinh}((\lambda_1-\lambda_2)w/2)&0& [\beta_{\lambda_1,\zeta}\eta_{\lambda_2}-\beta_{\lambda_2,\zeta}\eta_{\lambda_1}]\text{sinh}((\lambda_1+\lambda_2)w/2)&0\\
-\eta_{\lambda_1}\dfrac{\text{sinh}((\lambda_1-\lambda_2)w/2)}{\text{sinh}(\lambda_2w)}&0&\eta_{\lambda_1}\dfrac{\text{sinh}((\lambda_1+\lambda_2)w/2)}{\text{sinh}(\lambda_2w)}&\eta_{\lambda_2}
\end{pmatrix} \begin{pmatrix} A_1\\A_2\\A_{-1}\\A_{-2}
\end{pmatrix}=\begin{pmatrix} 0\\0\\0\\0\end{pmatrix}
\label{eqn:fundamentalsystem2}
\end{align*}
\end{widetext}

As in the Isolated edge solution, the quantities between brackets like $ [\beta_{-\lambda_1,\zeta}\eta_{\lambda_2}-\beta_{-\lambda_2,\zeta}\eta_{\lambda_1}]$ provides the proper dispersion relations for different solutions. Then for future convenience we will define:

\begin{equation}
\begin{cases} [\beta_{\lambda_1,\zeta}\eta_{\lambda_2}-\beta_{\lambda_2,\zeta}\eta_{\lambda_1}]=-\upsilon(\lambda_1-\lambda_2)\Delta E(\lambda_1,\lambda_2)\\
[\beta_{-\lambda_1,\zeta}\eta_{\lambda_2}-\beta_{-\lambda_2,\zeta}\eta_{\lambda_1}]=\upsilon(\lambda_1-\lambda_2)\Delta E(-\lambda_1,-\lambda_2)\\
[\beta_{\lambda_2,\zeta}\eta_{\lambda_1}-\beta_{-\lambda_1,\zeta}\eta_{\lambda_2}]=-\upsilon(\lambda_1+\lambda_2)\Delta E(-\lambda_1,\lambda_2)\\
[\beta_{-\lambda_2,\zeta}\eta_{\lambda_1}-\beta_{\lambda_1,\zeta}\eta_{\lambda_2}]=\upsilon(\lambda_1+\lambda_2)\Delta E(\lambda_1,-\lambda_2)
\end{cases}
\Delta E(\lambda_1,\lambda_2)=\theta_--E + c_{2-}[k_x^2+k_x\zeta(\lambda_1 + \lambda_2)+\lambda_1\lambda_2]
\end{equation}

\begin{widetext}
Then, the system reads:

\begin{align}
\begin{aligned}
\begin{pmatrix}
(\lambda_1-\lambda_2)\Delta E(-\lambda_1,-\lambda_2)\text{sinh}((\lambda_1+\lambda_2)w/2)& 0&(\lambda_1+\lambda_2)\Delta E(\lambda_1,-\lambda_2)\text{sinh}((\lambda_1-\lambda_2)w/2)&0\\ 
\eta_{\lambda_1,\zeta}\text{sinh}((\lambda_1+\lambda_2)w/2) & \eta_{\lambda_2,\zeta}\text{sinh}(\lambda_2w)&-\eta_{\lambda_1,\zeta}\text{sinh}((\lambda_1-\lambda_2)w/2)&0\\
-(\lambda_1+\lambda_2)\Delta E(-\lambda_1,\lambda_2)\text{sinh}((\lambda_1-\lambda_2)w/2)&0& -(\lambda_1-\lambda_2)\Delta E(\lambda_1,\lambda_2)\text{sinh}((\lambda_1+\lambda_2)w/2)&0\\
-\eta_{\lambda_1,\zeta}\text{sinh}((\lambda_1-\lambda_2)w/2)& 0&\eta_{\lambda_1,\zeta}\text{sinh}((\lambda_1+\lambda_2)w/2)&\eta_{\lambda_2,\zeta}\text{sinh}(\lambda_2w/2)
\end{pmatrix}\\ \times\begin{pmatrix} A_1\\A_2\\A_{-1}\\A_{-2}
\end{pmatrix}=\begin{pmatrix} 0\\0\\0\\0\end{pmatrix}
\end{aligned}
\label{eqn:fundamentalsystem3}
\end{align}
For a non-trivial solution in the coefficients $A_1, A_2, A_{-1}, A_{-2}$ we need the last determinant to be zero and we have the desired dispersion relation:
\begin{equation}
\begin{aligned}
(\lambda_1-\lambda_2)^2\Delta E(-\lambda_1,-\lambda_2)\Delta E(\lambda_1,\lambda_2)\text{sinh}^2((\lambda_1+\lambda_2)w/2)=(\lambda_1+\lambda_2)^2\Delta E(\lambda_1,-\lambda_2)\Delta E(-\lambda_1,\lambda_2)\text{sinh}^2((\lambda_1-\lambda_2)w/2)
\label{eqn:disp2}
\end{aligned}
\end{equation}

This is a quadratic equation in $E$ so, it can be explicitly obtained. Calling $g^{\pm}=(\lambda_1\pm\lambda_2)^2\text{sinh}^2((\lambda_1\mp\lambda_2)w/2)$:

\begin{equation*}
\begin{aligned}
g^-[(E-\theta_{-})+c_{2-}(k_x^2+\zeta k_x(\lambda_1+\lambda_2)+\lambda_1\lambda_2)][(E-\theta_{-})+c_{2-}(k_x^2-\zeta k_x(\lambda_1+\lambda_2)+\lambda_1\lambda_2)]=\\g^+[(E-\theta_{-})+c_{2-}(k_x^2+\zeta k_x(\lambda_1-\lambda_2)-\lambda_1\lambda_2)][(E-\theta_{-})+c_{2-}(k_x^2+\zeta k_x(\lambda_2-\lambda_1)-\lambda_1\lambda_2)]
\end{aligned}
\end{equation*}

 This is:
 
\begin{equation*}
\begin{aligned}
g^-[(E-\theta_{-}+c_{2-}k_x^2+c_{2-}\lambda_1\lambda_2)^2-c_{2-}^2\zeta^2k_x^2(\lambda_1+\lambda_2)^2]=g^+[(E-\theta_{-}+c_{2-}k_x^2-c_{2-}\lambda_1\lambda_2)^2-c_{2-}^2\zeta^2k_x^2(\lambda_1-\lambda_2)^2]=\\ =g^-[(E-\theta_{-})^2+(c_{2-}k_x^2+c_{2-}\lambda_1\lambda_2)^2+2(c_{2-}k_x^2+c_{2-}\lambda_1\lambda_2)(E-\theta_{-})-c_{2-}^2\zeta^2k_x^2(\lambda_1+\lambda_2)^2]=\\
=g^+[(E-\theta_{-})^2+(c_{2-}k_x^2-c_{2-}\lambda_1\lambda_2)^2+2(c_{2-}k_x^2-c_{2-}\lambda_1\lambda_2)(E-\theta_{-})-c_{2-}^2\zeta^2k_x^2(\lambda_1-\lambda_2)^2]
\end{aligned}
\end{equation*}

And then, simplifying we arrive at the desired equation:

\begin{equation*}
\begin{aligned}
(E-\theta_{-})^2(g^--g^+)+2c_{2-}[k_x^2(g^--g^+)+\lambda_1\lambda_2(g^-+g^+)](E-\theta_{-})+\\g^-[(c_{2-}k_x^2+c_{2-}\lambda_1\lambda_2)^2-c_{2-}\zeta^2k_x^2(\lambda_1+\lambda_2)^2]-g^+[(c_{2-}k_x^2-c_{2-}\lambda_1\lambda_2)^2-c_{2-}\zeta^2k_x^2(\lambda_1-\lambda_2)^2]=0
\end{aligned}
\end{equation*}

Rearranging terms:

\begin{equation*}
\begin{aligned}
(E-\theta_{-})^2(g^--g^+)+2c_{2-}[k_x^2(g^--g^+)+
\lambda_1\lambda_2(g^-+g^+)](E-\theta_{-})+\\
+(g^--g^+)c_{2-}^2[k_x^4+\lambda_1^2\lambda_2^2-\zeta^2k_x^2(\lambda_1^2+\lambda_2^2)]+2c_{2-}^2k_x^2(g^-+g^+)\lambda_1\lambda_2(1-\zeta^2)=0
\end{aligned}
\end{equation*}

Now, if we define $\varGamma=\dfrac{g^++g^-}{g^+-g^-}$ it is:

\begin{equation*}
\begin{aligned}
(E-\theta_{-})^2+2c_{2-}[k_x^2-\lambda_1\lambda_2\varGamma](E-\theta_{-})+c_{2-}^2[k_x^4+\lambda_1^2\lambda_2^2-\zeta^2k_x^2(\lambda_1^2+\lambda_2^2)-2k_x^2\varGamma\lambda_1\lambda_2(1-\zeta^2)]=0
\end{aligned}
\end{equation*}

And the two solutions are

\begin{equation}
E=\theta_-+c_{2-}\left[k_x^2-\lambda_1\lambda_2\varGamma\pm\sqrt{\zeta^2k_x^2(\lambda_1^2+\lambda_2^2-2\lambda_1\lambda_2\varGamma)+\lambda_1^2\lambda_2^2(\varGamma^2-1)}\right]
\label{eqn:disp3app}
\end{equation}
\end{widetext}

In the case $w\to\infty$ it is $\varGamma\to -1$ and we recover the Isolated edge solution (\ref{eqn:dispertionIE1app})\\
\\
In principle, equations (\ref{eqn:disp3app},\ref{eqn:rootsapp}), determines implicitly the Energy and the two roots $\lambda_{1,2}$. Through the first equation of system (\ref{eqn:fundamentalsystem3}) a relation between $A_1$ and $A_-1$ can be obtained, thus the wave function in (\ref{eqn:wavef2}) is then known, apart from normalization constant. 
Finally, we will derive another system of equations evaluating the Hamiltonian equation at the boundaries. This system is useful to find Energy independent quantities (and independent of the parameters $k_y, \theta_+, \theta_-$ in this model). The system is the equivalent of (\ref{eqn:trick2systemapp}) for the slab solution. Although not as useful as that, since it is much more complicated, it serves to compare relations between the slab and the isolated edge limit. So again, taking into account that $H\psi|_{\pm w/2}=0$:

\begin{align*}
\begin{pmatrix}
-c_{2-}\partial_y^2 & \upsilon \partial_y\\- \upsilon \partial_y   &  -c_{2+}\partial_x^2
\end{pmatrix}
\begin{pmatrix} \psi_{1, \zeta}\\\psi_{2, \zeta}
\end{pmatrix}\lvert_{\pm w/2}=\begin{pmatrix} 0\\0\end{pmatrix}
\end{align*}

Then, using the expression for the wave function (\ref{eqn:wavef3}), we have the following system:

\begin{widetext}
\begin{equation}
\begin{aligned}
\begin{pmatrix}
-c_{2-}\partial_y^2f^- &  -\upsilon \partial_yf^-\\  \upsilon \partial_yf^-&  -c_{2+}\partial_x^2f^- 
\end{pmatrix}A_{1}\Phi_{\lambda_1, \zeta}+\begin{pmatrix}
-c_{2-}\partial_x^2f^+ & \upsilon \partial_xf^+\\- \upsilon \partial_xf^+&  -c_{2+}\partial_x^2f^+ 
\end{pmatrix}A_{-1}\Phi_{-\lambda_1, \zeta}=\begin{pmatrix}0\\0\end{pmatrix}\\
\begin{pmatrix}
-c_{2-}\partial_x^2f^+ & -\upsilon \partial_xf^+\\  \upsilon \partial_xf^+&  -c_{2+}\partial_x^2f^+
\end{pmatrix}A_{1}\Phi_{\lambda_1, \zeta}+\begin{pmatrix}
-c_{2-}\partial_x^2f^- & \upsilon \partial_xf^-\\- \upsilon \partial_xf^-&  -c_{2+}\partial_x^2f^-
\end{pmatrix}A_{-1}\Phi_{-\lambda_1, \zeta}=\begin{pmatrix}0\\0\end{pmatrix}
\label{tricksystem}
\end{aligned}
\end{equation}

Where we have defined:

\begin{equation}
\begin{cases}\partial_y^2f^\pm=(\lambda_1^2-\lambda_2^2)e^{\pm\lambda_1w/2}\\\partial_yf^\pm=(\lambda_1-\lambda_2)e^{\pm\lambda_1w/2}-2\lambda_2\dfrac{\text{sinh}((\lambda_1-\lambda_2)w/2)}{\text{sinh}(\lambda_2w)}e^{\mp\lambda_2w/2}
\end{cases}
\label{trickcases}
\end{equation}

After some algebraic manipulations this system reads:

\begin{align*}
\begin{pmatrix}
& -c_{2-}\partial_y^2f^- \, \upsilon \partial_yf^- \,  -c_{2-}\partial_y^2f^+ \,  \upsilon \partial_yf^+\\
& 0 \, -[c_{2-}c_{2+}(\partial_y^2f^-)^2+\upsilon^2(\partial_yf^-)^2] \, -\upsilon c_{2-}(\partial_y f^+\partial_y^2f^- + \partial_yf^-\partial_y^2f^+) \, -c_{2-}c_{2+}(\partial_y^2f^+\partial_y^2f^-)+\upsilon^2(\partial_yf^+\partial_yf^-)\\
& -c_{2-}\partial_y^2f^+ \, -\upsilon \partial_yf^+ \,  -c_{2-}\partial_y^2f^- \, \upsilon \partial_yf^-\\
& 0 \, -[c_{2-}c_{2+}(\partial_x^2f^+)^2+\upsilon^2(\partial_yf^+)^2] \, -\upsilon c_{2-}(\partial_y f^-\partial_y^2f^+ + \partial_yf^+\partial_y^2f^-) \, -c_{2-}c_{2+}(\partial_y^2f^+\partial_y^2f^-)+\upsilon^2(\partial_yf^+\partial_yf^-)
\end{pmatrix}\times
\end{align*}

\begin{flushleft}
	$\times\begin{pmatrix}
	A_{1}\Phi_{\lambda_1, \zeta}\\A_{-1}\Phi_{-\lambda_1, \zeta}
	\end{pmatrix}=\begin{pmatrix}0\\0\\0\\0\end{pmatrix}$
\end{flushleft}
\end{widetext}

As always, the determinant must be zero to have a non-trivial solution, and in this system this is so, only if the second and the fourth row are the same. This is: $c_{2-}c_{2+}(\partial_x^2f^+)^2+\upsilon^2(\partial_xf^+)=c_{2-}c_{2+}(\partial_x^2f^-)^2+\upsilon^2(\partial_xf^-)$. Substitution of (\ref{trickcases}) gives an implicit closed equation for the two roots $\lambda_{1,2}$ that depends only in the parameters $c_{2+}, c_{2-}, \upsilon$: $(\lambda_1-\lambda_2)^2(c_{2-}c_{2+}(\lambda_1+\lambda_2)^2+\upsilon^2)=\dfrac{4\upsilon^2\text{sinh}^2((\lambda_1-\lambda_2)w/2)}{\text{sinh}(\lambda_1w)\text{sinh}(\lambda_2w)}\lambda_1\lambda_2$. This can be related with the $\varGamma$ quantity that appears in the dispersion relation (\ref{eqn:disp3}), so we arrive at the final expression:

\begin{equation}
2\lambda_1\lambda_2c_{2-}c_{2+}(\varGamma+1)=\upsilon^2+c_{2-}c_{2+}(\lambda_1+\lambda_2)^2
\label{trickequationapp}
\end{equation}

Then, it is clear that when $\varGamma\xrightarrow[w\to\infty]{} -1$ result (\ref{trickequationapp}) is reduced to condition (\ref{eqn:tricksol12sysapp}) for the two roots in the isolated edge solution. Then in that case, through the system (\ref{tricksystem}) we recover the position-independent expression for the two component spinor.

Systems (\ref{eqn:fundamentalsystem3}),(\ref{tricksystem}) and the info obtained from their determinants (\ref{eqn:disp2}, \ref{trickequationapp}), together with the expression (\ref{eqn:wavef2app}) for the wave function, and relation (\ref{eqn:rootsapp}) for the two roots are the fundamental equations we will use to derive all quantities of interest with slab boundary conditions.
\end{document}